\documentclass[12pt,preprint]{aastex}

\newcommand{\um}{\ensuremath{\mathrm{\,\mu{}m}}}
\newcommand{\msun}{\ensuremath{\mathit{\,M_\odot}}}
\newcommand{\kps}{\ensuremath{\mathrm{\,km\,s^{-1}}}}

\newcommand{\period}{\ensuremath{\mathit{P}}}
\newcommand{\e}{\ensuremath{\mathit{e}}}
\newcommand{\Tknot}{\ensuremath{\mathit{T}_0}}
\newcommand{\w}{\ensuremath{\mathit{\omega}}}
\newcommand{\qmin}{\ensuremath{\mathit{q}_{\mathrm{min}}}}
\newcommand{\q}{\ensuremath{\mathit{q}}}
\newcommand{\Kone}{\ensuremath{\mathit{K}_1}}
\newcommand{\Ktwo}{\ensuremath{\mathit{K}_2}}
\newcommand{\g}{\ensuremath{\mathit{\gamma}}}
\newcommand{\fmass}{\ensuremath{\mathit{f(M)}}}
\newcommand{\vrotsini}{\ensuremath{\mathit{v}_{\mathrm{rot}}\sin i}}
\newcommand{\Mone}{\ensuremath{\mathit{M}_1}}
\newcommand{\Mtwo}{\ensuremath{\mathit{M}_2}}

\newcommand{\Vtwo}{\ensuremath{\mathit{V}_2}}

\newcommand{\phnn}{\phantom{00}}
\newcommand{\phnnn}{\phantom{000}}
\newcommand{\phnnnn}{\phantom{0000}}
\newcommand{\phnnnnn}{\phantom{00000}}
\newcommand{\phnnnnnn}{\phantom{000000}}
\newcommand{\phdn}{\phantom{.0}}
\newcommand{\phdnn}{\phantom{.00}}
\newcommand{\phdnnn}{\phantom{.000}}
\newcommand{\phdnnnn}{\phantom{.0000}}

\newcommand{\phdnnnnnn}{\phantom{.000000}}
\newcommand{\php}{\phantom{+}}
\newcommand{\phnp}{\phantom{0+}}

\begin{document}

\title{The Detection of Low Mass Companions in Hyades Cluster
  Spectroscopic Binary Stars}

\author{Chad F.\ Bender\altaffilmark{1,2} and Michal
  Simon\altaffilmark{2}}

\altaffiltext{1}{Naval Research Laboratory, Remote Sensing Division:
  Code 7211, 4555 Overlook Ave.\ SW, Washington, DC 20375;
  chad.bender@nrl.navy.mil}

\altaffiltext{2}{Department of Physics and Astronomy, Stony Brook
  University, Stony Brook, NY 11794-3800; michal.simon@sunysb.edu}

\shortauthors{Bender and Simon}
\shorttitle{Hyades Spectroscopic Binary Stars}
\slugcomment{Accepted for publication in The Astrophysical Journal}

\begin{abstract}

  We have observed a large sample of spectroscopic binary stars in the
  Hyades Cluster, using high resolution infrared spectroscopy to
  detect low mass companions.  We combine our double-lined infrared
  measurements with well constrained orbital parameters from visible
  light single-lined observations to derive dynamical mass ratios.
  Using these results, along with photometry and theoretical
  mass-luminosity relationships, we estimate the masses of the
  individual components in our binaries.  In this paper we present
  double-lined solutions for 25 binaries in our sample, with mass
  ratios from $\sim0.1-0.8$.  This corresponds to secondary masses as
  small as $\sim0.15\msun$.  We include here our preliminary detection
  of the companion to vB~142, with a very small mass ratio of
  $\q=0.06\pm0.04$; this indicates that the companion may be a brown
  dwarf.  This paper is an initial step in a program to produce
  distributions of mass ratio and secondary mass for Hyades cluster
  binaries with a wide range of periods, in order to better understand
  binary star formation.  As such, our emphasis is on measuring these
  distributions, not on measuring precise orbital parameters for
  individual binaries.
\end{abstract}


\section{\sc Introduction\label{intro}}

Observations of spectroscopic binary stars provide dynamical
measurements of stellar mass and binary mass ratio that are important
inputs to theories of binary star formation \citep{bonnell2003,
  tohline2002, clarke2001}.  Measuring the radial velocity versus
orbital phase for a single-lined spectroscopic binary (SB1) yields
orbital parameters and the mass function, \fmass, in which the mass
ratio, $\q=\Mtwo/\Mone$, is inseparable from the orbital inclination.
For an ensemble of SB1s, a statistical approach can be used to derive
the distribution of mass ratios \citep[e.g.,][]{mazeh1992}.  If,
however, a binary is observed as a double-lined system (SB2), then the
dynamical mass ratio follows directly from the radial velocity
measurements \citep{mazeh2002}.  With an estimate of the primary mass,
say based on the spectral type, the mass ratio gives the secondary
mass, and in the case of an ensemble of SB2s, the secondary mass
distribution.

Obtaining a purely dynamical mass ratio distribution using SB2s is
challenging: a binary with a mass ratio of much less than one has a
small flux ratio, $\alpha=\mathit{F}_2/\mathit{F}_1$, so the secondary
component is difficult to detect.  This flux ratio problem is
particularly inhibitive at visible wavelengths where much of the long
term monitoring of spectroscopic binaries has occurred.  For this
reason, most identified SBs are SB1s, and there is a strong selection
effect favoring the detection of SB2s with \q{} near 1. A binary
composed of main sequence stars with unequal masses will have an
$\alpha$ that increases towards longer wavelengths, making the
companion easier to detect using infrared spectroscopy.  Such
observations have allowed the measurement of SB2s with \q's as small
as $\sim0.1-0.2$ \citep[e.g.,][]{mazeh2002,prato2002}.

\citet{duquennoy1991} examined a sample of 164 nearby SB1s, SB2s,
visual binaries, and common proper motion pairs, with F- and G-type
primary stars, and found that the mass ratio distribution for medium
to long period binaries increases towards small mass ratios.
\citet{goldberg2003} used a well defined sample of 129 SB1s and SB2s
from the Carney-Latham sample of high proper motion stars
\citep{carney1994} to derive a bimodal mass ratio distribution with
peaks at $\q\sim0.2$ and $\q\sim0.8$.  They found somewhat different
distributions for halo versus disk stars, and for binaries with
primary masses greater than and less than 0.67\msun.  By working in
the infrared, \citet{mazeh2003} observed 32 binaries from the
Carney-Latham sample as SB2s, with mass ratios as small as $\sim$0.20.
When combined with SB2s from visible light spectroscopy, they found a
mass ratio distribution that is approximately flat from $\q\sim0.3 -
1.0$. 

The samples studied by \citet{goldberg2003}, \citet{mazeh2003}, and,
for \q's near 1 by \citet{lucy2006}, represent averages of star
formation outcomes in the solar neighborhood over billions of years.
Most star formation occurs in localized regions within molecular
clouds.  Studies of the binary population in diverse star forming
regions can isolate the physical parameters that determine the
properties of the binaries formed.  For example, the competing effects
of fragmentation, dynamical interactions, and accretion are not fully
understood \citep{goodwin2007, ballesteros2007, whitworth2007}.
Observations of binaries in open clusters can clarify the important
processes because they share a common star formation origin and
history.  Ideally, such observations would be carried out on young
clusters that have an easily identifiable membership and still retain
their most massive members.  However, it is difficult to measure the
orbital velocities of the youngest stars because they tend to have
high rotation velocities.  Stars in older open clusters still retain
much of their cluster identity yet are rotationally spun down.

The Hyades is one such nearby open cluster: it is well studied and
presents an excellent laboratory for measuring a well defined sample
of binary stars in order to investigate binary star formation.  The
cluster has an age of $\sim650$\,Myr \citep{lebreton2001}, old enough
that stars later than spectral type F3 have slowed to $v\sin i \la
25\kps$ \citep{kraft1965}, and G-type and later to $v\sin i\la10\kps$
\citep{stauffer1987}.  Most mid K-type and earlier cluster members
have individual \emph{Hipparcos} distances, and with a mean cluster
distance of $\sim46$\,pc \citep{perryman1998} the known binaries are
bright.  Additionally, with metallicity [Fe/H]=0.14
\citep{lebreton2001} the spectral lines of Hyades members are deep and
well suited to spectroscopic analysis.

Radial velocity surveys of the Hyades \citep[e.g.,][]{wilson1948,
  kraft1965, detweiler1984, stefanik1985,
  griffin1978,griffin1981,griffin1985, griffin1988} have identified a
sufficient number of SBs to facilitate a statistical analysis of their
physical properties. \citet{patience1998} carried out a study of
Hyades ``visual binaries'' using speckle interferometry.  These wide
binaries have periods longer than most SBs and several are the wide
components of hierarchical triples with known SBs.  From their
observations, \citet{patience1998} derived a photometric mass ratio
distribution for long period systems that appears to increase towards
small \q.

At the Harvard-Smithsonian Center for Astrophysics (CfA), Robert
P. Stefanik (RPS) and David W. Latham (DWL) have monitored Hyades
members for SBs by visible light spectroscopy since 1979
\citep[e.g.,][]{latham1982,stefanik1985}.  Their observations have
refined the cluster membership and yielded precise parameters for many
SB1s and SB2s.  In 2003, the current authors began a collaboration
with the CfA group to extend, through infrared observations, the
detection of binary companions to smaller masses than possible by
visible light spectroscopy alone.  To this end, RPS and DWL made
available to us their unpublished parameters for many Hyades SB1s.  In
this paper we present infrared SB2 detections that span the range of
mass ratios from $\q\sim0.1$ to $\q\sim0.9$.  We verified these
results using the sample of Hyades SB1s available in the literature
\citep[e.g.,][]{sanford1921,griffin1978,griffin1981,griffin1985}.  When
combined with available surveys (both present and future) of Hyades
binaries, our SB2 detections will enable better determinations of the
distribution of secondary masses.  In \S\ref{sample}, we present our
Hyades binary sample; \S\ref{infrared} describes our infrared
observations and velocity measurement techniques; \S\ref{sb2solutions}
includes the SB1 and SB2 solutions for our sample binaries;
\S\ref{discussion} discusses the quality of our SB2 solutions and
derives the secondary masses; and \S\ref{summary} presents a brief
summary and comments on our future papers.

\section{\sc The Infrared SB2 Sample \label{sample}}

Because the Hyades cluster is nearby, it covers a large angular extent
on the sky, so determining the cluster membership is difficult.
Considerable effort has been applied to identify true Hyades members
using measurements of proper motion, radial velocity, parallax, and
color-magnitude relationships \citep[e.g.,][]{vanbueren1952,
  vanaltena1969, hanson1975, griffin1988, reid1992, perryman1998,
  debruijne2001}.  Nevertheless, these surveys contain many candidates
that are non-members or have uncertain membership status.  In
particular the membership of SBs is often ambiguous until their
center-of-mass velocities can be calculated from orbital solutions.
The multi-decade long campaign at CfA to monitor candidate Hyades
binaries and multiple systems (\S\ref{intro}) has, for many systems,
resulted in precise measurements of both the orbital parameters (the
period, \period, the eccentricity, \e, the time of periastron passage,
\Tknot, the longitude of periastron, \w, the semi-major amplitude of
the primary, \Kone, and the center-of-mass velocity, \g) and the mass
function, \fmass.

\begin{figure}
  \epsscale{1.0}
  \plotone{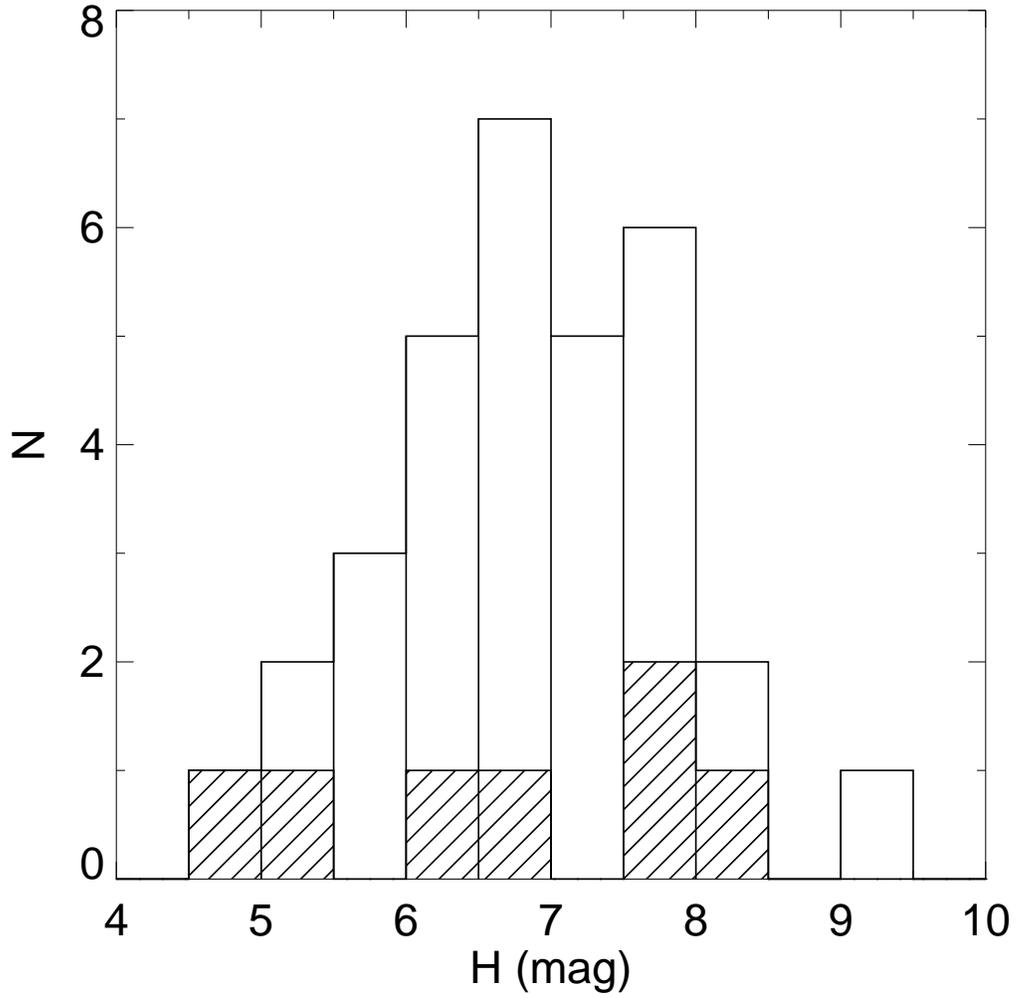}
  \caption{The distribution of 2MASS \emph{H}-magnitudes for the
    infrared sample binaries.  The \emph{hashed} region indicates
    binaries for which we did not detected a companion
    (\S\ref{sb2nondetections}).}
  \label{fig:hmag}
\end{figure}

In 2003, we used the preliminary CfA SB1 solutions to select a sample
of 32 binaries suitable for further observations in the infrared as
SB2s.  We will refer to this as the \emph{infrared sample}, to
distinguish from the entire CfA sample.  All infrared sample members
are confirmed as cluster members through the methods listed above.  To
avoid early type stars with large rotational velocities
(\S\ref{intro}) we restricted the sample to targets with primaries of
spectral type F and later.  Despite this, however, five targets do
have $V_{rot}\ga25\kps$ and required a modified analysis
(\S\ref{irrapidrotators}).  Most of the infrared observations were
obtained with the CSHELL spectrometer at the IRTF, which, for Hyades
members, has an effective magnitude limit of \emph{H}$\sim$9 in a 90
minute integration; this corresponds to targets with early-M primary
stars, and serves as the faint cutoff for our sample.  We intended the
infrared observing campaign to last at most a few years, during which
the orbital phase of systems with long periods would not change
significantly.  Using \fmass{} from the SB1 solutions, an estimate of
the primary mass based on spectral type, and letting $i=\pi/2$, we
derived the minimum mass ratio, \qmin, allowed for each binary.  From
\qmin{} we predicted the velocity separation of long period systems,
and excluded those with separations of only a few \kps{} or smaller.
Long period systems with large predicted velocity separations were
retained, with the expectation that observations at multiple phases
would not be possible.  Table~\ref{table:irsample} lists the infrared
sample.  Columns 1--5 give the name and HD number of the primary star,
the J2000 coordinates, and the primary spectral type as listed in the
SIMBAD database and confirmed with the analysis described in
\S\ref{infrared}.  Columns 6 and 7 give \emph{V} magnitude
(\emph{Hipparcos} where available and SIMBAD otherwise) and \emph{H}
magnitude (2MASS), respectively.  Figure~\ref{fig:hmag} shows the
distribution of \emph{H}-magnitude, from 2MASS, for the infrared
sample.  The hashed regions indicate sample members for which we did
not detect a companion. (see \S\ref{sb2nondetections}).

\section{\sc Infrared Spectroscopy and Double-Lined
  Measurements\label{infrared}}

\subsection{\sc Observations\label{irobs}}

We observed the infrared sample at the NASA Infrared Telescope
Facility (IRTF) with CSHELL, the facility near-infrared echelle
spectrograph \citep{greene1993}; Table~\ref{table:obslog} contains a
log of the observations.  CSHELL uses a 256x256 pixel InSB detector
with a plate scale of $0\farcs2\,\mathrm{pixel}^{-1}$, and operates
under natural seeing at the IRTF.  Echelle orders are isolated with an
order sorting circular variable filter.  All of the observations used
a single grating setting centered at 1.5548\um{} because this spectral
region contains several deep lines, a requirement for precisely
measuring radial velocities.  We used a $0\farcs5$ slit which provided
a spectral resolution of $\sim30,\!000$.  The spectra we obtained have
a free spectral range of $\sim40$\AA.

We observed each binary as a series of short integrations, from 60 to
300\,s in duration, nodding the telescope $\sim10\arcsec$ along the
slit between integrations in an ABBA pattern.  By differencing each
set of A and B frames we removed contributions from detector bias,
dark current, and sky background.  We took a series of flat and dark
frames at the beginning of each observing session, which were median
filtered and normalized to provide a flat field correction for each
target ``A-B'' image. To measure the dispersion solution, we observed
Ar and Kr arc-lamp spectra from CSHELL's internal lamps.  We extracted
each ``A-B'' image using a custom optimal extraction code written in
IDL, patterned after the algorithms described by \citet{piskunov2002},
and averaged the set of resulting spectra for each target.  We
adjusted the total integration time for each target to provide spectra
with a signal-to-noise of $\mathrm{S/N}\sim100$, although we relaxed
this condition for targets with a large \qmin, and also for the
faintest targets to avoid very long integrations that were an
inefficient use of telescope time.  Observations in October 2003
suffered from poor weather conditions and did not result in any usable
spectra.  During the 2005 and 2006 observations the CSHELL detector
experienced brief, uncontrolled temperature warm-ups that increased
the thermal background by as much as two orders of magnitude over the
nominal few tens of ADUs, affecting $\sim20\%$ of the integrations.
Spectra obtained under these circumstances have decreased S/N, but
most are still usable.

The CSHELL observations from 2004 suggested that high S/N observations
of H69 and vB~142 with a larger telescope would be particularly
useful.  H69, with $H\sim9.1$ mag, is a faint target for the IRTF.
The SB1 solution for vB~142 yields a mass function with a very small
\qmin{} of 0.04, and the 2004 October 1 SB2 measurement indicated that
\q{} was indeed small and the companion possibly a brown dwarf.  We
subsequently observed these binaries on 2005 February 22 at the W.  M.
Keck Observatory with NIRSPEC, the facility near-infrared spectrograph
\citep{mclean2000}.  We also observed vB~43, which at that time was
nearing a phase with a large velocity separation.  These observations
are included in Table~\ref{table:obslog}.  We used NIRSPEC in its
cross-dispersed echelle mode with the 2-pixel slit, which provided a
spectral resolution of $\sim31,\!000$.  Each target was observed as a
sequence of ABBA integrations, using a nod of $\sim10\arcsec$.  A
sequence of flat and dark frames was obtained at the beginning of the
night and median filtered.  Night sky OH emission lines, identified
from the catalog of \citet{rousselot2000}, provided a simultaneous
wavelength reference to determine the dispersion solution. We
extracted the spectra in IDL using REDSPEC\footnote{See
  http://www2.keck.hawaii.edu/inst/nirspec/redspec/}, the NIRSPEC
facility software package.  In addition to normal image processing,
REDSPEC rectifies the images in both the spatial and spectral
dimensions to extract each order.  The extracted nod pairs were
averaged for each target.  We worked only with NIRSPEC order 49,
centered at $\sim1.56\um$ and spanning $\sim220$\AA{}, because it is
nearly completely free of telluric absorption; the other orders
provided by our grating setting are contaminated to varying degrees
\citep[see][Fig.~1]{bender2005}.

\subsection{\sc Template Spectra\label{irtemplates}}

We measure the radial velocities of a binary spectrum with reference
to template stellar spectra having spectral type, metallicity, and
rotational velocity similar to the binary components.  We have two
libraries of observed single star templates.  The first was obtained
with NIRSPEC and covers spectral types from G0 through M9 \citep[][Fig
3\,--\,6]{bender2005}.  The template velocities are tied to the
\citet{nidever2002} reference frame; this procedure and the individual
template velocities are reported in \citet{simon2006}.  The second
library was obtained with CSHELL and covers spectral types from G0
through M1 \citep[][Fig 2]{mazeh2002}.  This library is considerably
more sparse in spectral type compared with the NIRSPEC library and
some CSHELL templates have a low S/N.  Both template libraries are
comprised of stars with small rotational velocities.  We used a
non-linear limb darkening model \citep{claret2000} to rotationally
broaden the templates as needed, following the procedure outlined by
\citet{gray1992}.  In the analysis presented here, we relied primarily
on the NIRSPEC templates because of their high quality, but also
utilized the CSHELL templates where they matched the spectral types of
the targets.

\subsection{\sc Radial Velocity Measurements\label{irccorr}}

We analyzed the infrared spectra with a two-dimensional correlation
algorithm similar to the TODCOR algorithm described by
\citet{zucker1994}.  For each infrared observation, we used the SB1
parameters to calculate the orbital phase and the radial velocity of
the primary.  We then estimated the range of possible secondary
velocities at that phase, corresponding to \q{} ranging from \qmin{}
to 1, with some allowance for the uncertainty in the primary mass
estimate (\S\ref{sample}).  We analyzed each target with two or three
templates that matched the known characteristics of the primary and a
range of four to eight plausible secondary templates, while leaving
the flux ratio unconstrained.  This resulted in sets of measured
velocities and flux ratios for each combination of templates, which we
averaged to obtain the final measured velocities reported in
Table~\ref{table:obslog}. We did not detect the companion for seven of
the binaries in our sample, and for three others were successful for
only part of our observations.  We indicate these observations in
Table~\ref{table:obslog} as missing data, and address these systems in
detail in \S\ref{sb2nondetections}.

We estimated the uncertainties of both the primary and secondary
velocity measurements as a combination of two factors: (1) the
velocity uncertainty of the templates and (2) the uncertainty in
measuring the center of the correlation peak due to noise in the
target spectrum.  We took the contribution from the templates simply
as the variance of the velocities measured over the set of template
pairs.  We measured the contribution from noise in each target
spectrum as follows. First, we created a model binary from a pair of
templates used in the target spectrum analysis, with the same velocity
separation and flux ratio as measured in the target.  Then, we added
random noise to the model and analyzed it using the correlation
algorithm with the same pair of templates.  We adjusted the amplitude
of the noise until the peak correlation from the model equaled the
peak correlation measured in the target spectrum.  We then generated
100 copies of this model, each using a unique noise vector at the
prescribed amplitude.  Analyzing these models resulted in
distributions of primary and secondary velocities, which we fit 
as normal distributions.  We repeated this with four different
template pairs, and took the average of the distribution widths to be
the uncertainty contributed by noise in that target spectrum. By
combining this average in quadrature with the uncertainty due to the
templates, we arrived at our best estimate of the uncertainty for an
individual velocity measurement.  We used this procedure to
empirically determine the uncertainties on both our primary and
secondary velocity measurements for each of our observed target
spectra.

The relative contribution to the total uncertainties from (1) and (2)
varied for each target spectrum.  Templates with later spectral type
have a worse velocity precision that those with an earlier spectral
type \citep{bender2006}. So, for example, an analysis using M-star
templates would have a large portion of the total uncertainty
contributed by the templates.  Conversely, a target spectrum with low
S/N, often the case for the fainter binaries in our sample, would
derive most of its total uncertainty from the measurement of the
correlation peak.  An additional source of noise that has effected
previous correlation analyses \citep[e.g.,][]{bender2005} comes from a
mismatch between the target primary and the template primary.  If
these spectra have different physical properties, particularly the
metallicity of various species, then the precision of both the
measured velocities and flux ratio is effected.  We examined this on
several of our target spectra where the velocity uncertainties
obtained using the above procedure seemed inadequate (see
\S\ref{sb2solutions}).  We reanalyzed the model binaries from (2)
using alternate template spectra of similar spectral type, thereby
simulating potential mismatch between the templates and the target
spectra.  We found that the increase the velocity uncertainty over
that obtained in (2) was negligible.  This is likely due to the small
spectral range of our CSHELL templates, which limits the possibility
of metallicity mismatch.  Consequently, the velocity uncertainties
reported in Table~\ref{table:obslog} do not include any contribution
from this effect.

\begin{figure}
  \epsscale{1.0}
  \plotone{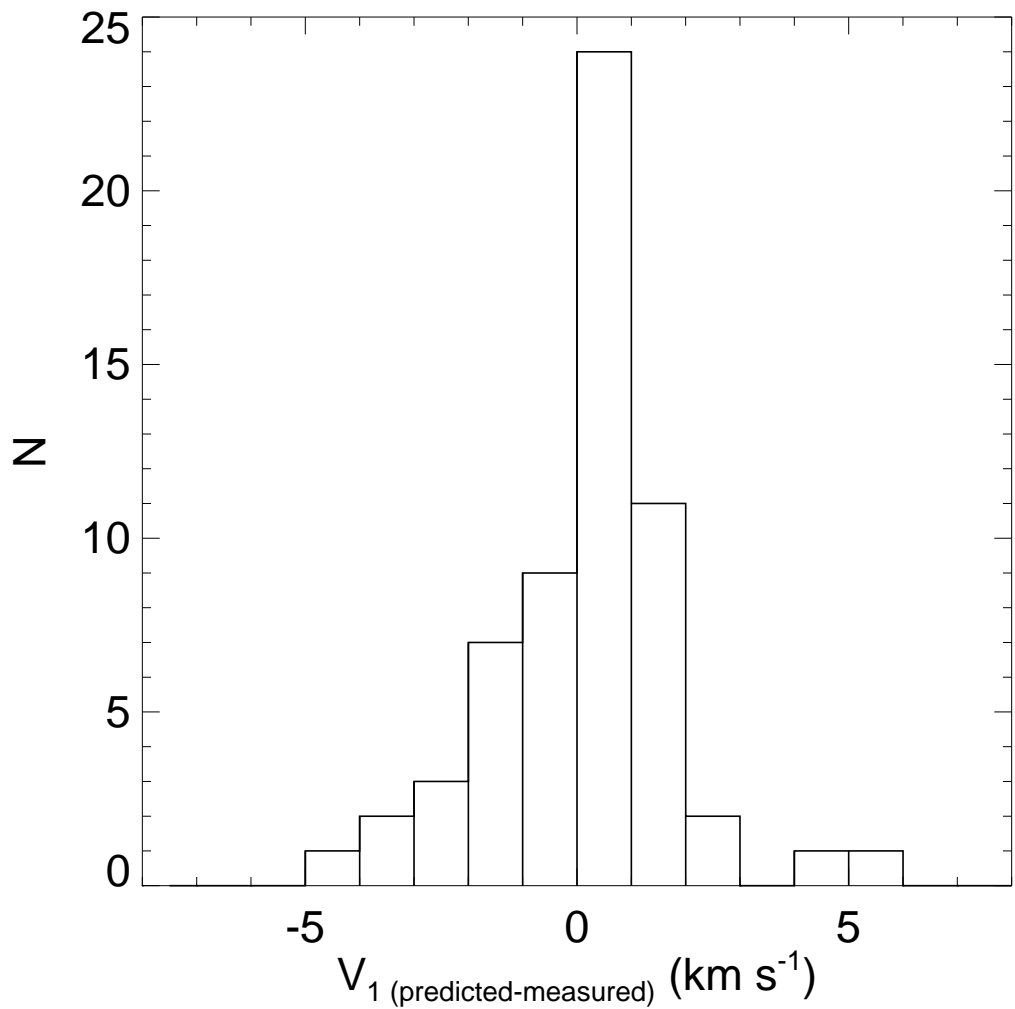}
  \caption{The distribution of the difference between the primary
    velocity predicted from the visible SB1 solutions and that
    measured from the infrared spectra.  The outlying measurements with
    large velocity difference are attributable to spectra with low
    S/N.  The distribution has a mean value of $\sim0.3\kps$, and is
    fit by a Gaussian with with $\sim0.9\kps$, indicating that the two
    reference frames agree to better than the infrared measurement
    uncertainties.}
  \label{fig:vonediff}
\end{figure}

To verify that our visible and infrared velocity reference frames were
in good agreement, we compared the primary velocities measured in the
infrared spectra with those predicted from the visible light SB1
solutions.  Figure~\ref{fig:vonediff} shows the distribution of the
difference between the predicted velocities and the measured
velocities.  The distribution has a mean of $\sim0.3\kps$ and is fit
by a Gaussian with a width of $\sim0.9\kps$.  This difference is
smaller than our infrared velocity precision \citep{simon2006} and
demonstrates that the two reference frames are indeed consistent to
within the measurement uncertainties.  The samples in
Figure~\ref{fig:vonediff} with a velocity difference larger than
$\sim5\kps$ are attributable to infrared spectra with a low S/N.

\subsection{\sc Rapid Rotators\label{irrapidrotators}}

The targets vB~8, vB~30, vB~68, vB~77, and BD+02~1102 have F-type
primary stars that are rapidly rotating, with \vrotsini{} from
$\sim40- 100\kps$, and so the spectral lines of these stars are
strongly rotationally broadened. The limited spectral range of the
CSHELL spectra did not contain enough strong features to measure the
primary velocities in the infrared spectra of these binaries.  We
considered, however, that lower mass companions will have spun down
and might therefore be sensitive to a correlation analysis.  To
examine this we modified our correlation analysis to use an artificial
primary template comprised of a featureless, flat spectrum.  We
analyzed each target spectrum with a range of plausible secondary
templates, allowing the flux ratio to vary, to measure the secondary
velocity.  This procedure was successful for vB~68, vB~77, and
BD+02~1102, and the velocities we measured for these systems are
included in Table~\ref{table:obslog}, along with a note indicating the
altered analysis.  The associated uncertainties were modeled using the
procedure described in \S\ref{irccorr} and the artificial primary
template described above.  We did not detect the secondaries for vB~8
or vB~30; we discuss both systems in \S\ref{sb2nondetections}.

\subsection{\sc Triple Systems\label{irtriples}}

Four binaries in our infrared sample, L20, vB~102, vB~151, and vB~40,
are the inner pair of hierarchical triple systems.  Speckle imaging
from \citet{patience1998} identified the wide companions in these
systems with orbital periods from a few tens to a few hundreds of
years, assuming circular orbits.  We briefly describe how the presence
of a third spectrum in our observations affected the analysis.

The wide companion to L20 has similar brightness to the primary of the
inner binary. Its spectrum is clearly apparent in our infrared
observations, with a radial velocity of $\sim32\kps$.  When left
unaccounted for, this contribution limited the precision of our radial
velocity and flux ratio measurements for the inner binary.  To correct
for this, we used a one-dimensional correlation to determine that the
wide companion spectrum was well matched with a K5 type template and
contributed about half of the total flux.  We then used the template
spectrum to subtract out this contribution prior to the normal
two-dimensional cross-correlation step.  In this manner, we increased
the precision of our inner binary velocity measurements and decreased
the uncertainty on \Ktwo{} by a factor of two over that obtained when
not accounting for the wide companion.

Using the component masses derived by \citet{patience1998} and the
\citet[][BCAH]{baraffe1998} stellar models, we estimate that the wide
companion to vB~102 contributes only $\sim10\%$ of the total flux at
1.6\um{}.  Additionally, its position angle \citep{patience1998}
placed it near the edge or outside of the CSHELL slit during our
observations.  If the outer binary is in a circular orbit, its period
is $\sim30$ years, and the velocity separation from the inner binary
\g-velocity is only a few \kps.  Additionally, the radial velocity of
the inner binary primary was only a few \kps{} from the \g-velocity
during our observations.  Consequently, any contamination in our
spectra from the outer companion was masked by the primary and we were
unable to detect it; we therefore solved vB~102 as a normal SB2.

The wide companions of vB~151 and vB~40 have large angular separations
and were at position angles such that they fell outside of the CSHELL
slit during our observations.  Even had they fallen within the slit,
their contributions to the total flux are small at 1.6\um, $\sim10\%$
and $\sim2\%$, respectively.  We did not detect any contribution from
the wide companion in either system, and concluded that our SB2
correlation analyses of the inner binaries were unaffected.

\section{\sc Infrared SB2 Solutions\label{sb2solutions}}

\subsection{\sc C\rm{}f\sc{}A SB1 Parameters}
With the permission of RPS and DWL, we list in columns 2--7 of
Table~\ref{table:orbitparams} the CfA orbital parameters for the 25
SB1s that we succeeded in turning into SB2s with infrared detections;
the seven sample members for which we did not detect the secondary
(\S\ref{irccorr}) are not included.  The CfA parameters are based on
preliminary orbital solutions that may change slightly when the final
results are published, but these changes are expected to be
insignificant for our present purposes because the uncertainties in
our final mass ratios are dominated by the infrared velocities for the
secondaries.

We have from one to three infrared spectra for most our binaries, and
none were observed on more than five occasions.  This contrasts with
the several tens or more visible light observations for each system.
Consequently, the primary velocities that we measured in the infrared
do not further improve the precision of the SB1 parameters.  To solve
each system as an SB2, we took, without modification, the SB1
parameters from Table~\ref{table:orbitparams}, and used a
least-squares fitting routine to solve the infrared \Vtwo{}
measurements for \Ktwo.  For several of our binaries (vB 62, vB 69,
H532, vB102, vB 142, vB 121, and vB 151), this procedure yielded
reduced chi-squared, $\chi^2_{\nu}$, much greater than one, indicating
that the corresponding \Vtwo{} uncertainties (\S3.3) are
underestimated.  Assuming that the missing uncertainty contribution in
\Vtwo{} is normally distributed, we account for it by scaling the
measured \Ktwo{} uncertainty by $\sqrt{\chi^2_{\nu}}$.  Because we do
not understand the source of this additional velocity uncertainty, we
choose to retain in Table~\ref{table:obslog} the uncertainties derived
in \S\ref{irccorr}.  To maintain consistent results throughout our
sample, we applied this correction to all of the binaries; column 8 of
Table~\ref{table:orbitparams} lists the derived \Ktwo{} along with the
adjusted uncertainties.

We list in column 9 of Table~\ref{table:orbitparams} the mass ratio
for each system, calculated from \Kone{} and \Ktwo{} as
$\q=\Kone/\Ktwo$.  We verified \q{} for the seven SB2s with large
$\chi^2_{\nu}$ using the technique of \citet{wilson1941}; in each
case, these results agree to within $1\sigma$ of those from the
least-squares fitting.  The uncertainty we report for \q{} includes
the unlisted uncertainty on \Kone{} from the preliminary CfA SB1
solution.  This contribution, however, is generally negligible when
compared with the large uncertainties on \Ktwo.  For example, \Kone{}
for vB~9 is determined to $\sim6\%$, which contributes only $\sim10\%$
of the total uncertainty on \q; most of the infrared sample members
have \Kone{} determined much more precisely.  L20 has \Ktwo{}
determined to better than $\sim4\%$ and is one of our most precisely
measured infrared systems; \Kone{} here contributes only $\sim1\%$ to
the total uncertainty on \q.  Columns 10--13 in
Table~\ref{table:orbitparams} give the semi-major axes and component
masses, combined with the unknown $\sin i$.

\begin{figure}
  \epsscale{0.8}
  \plotone{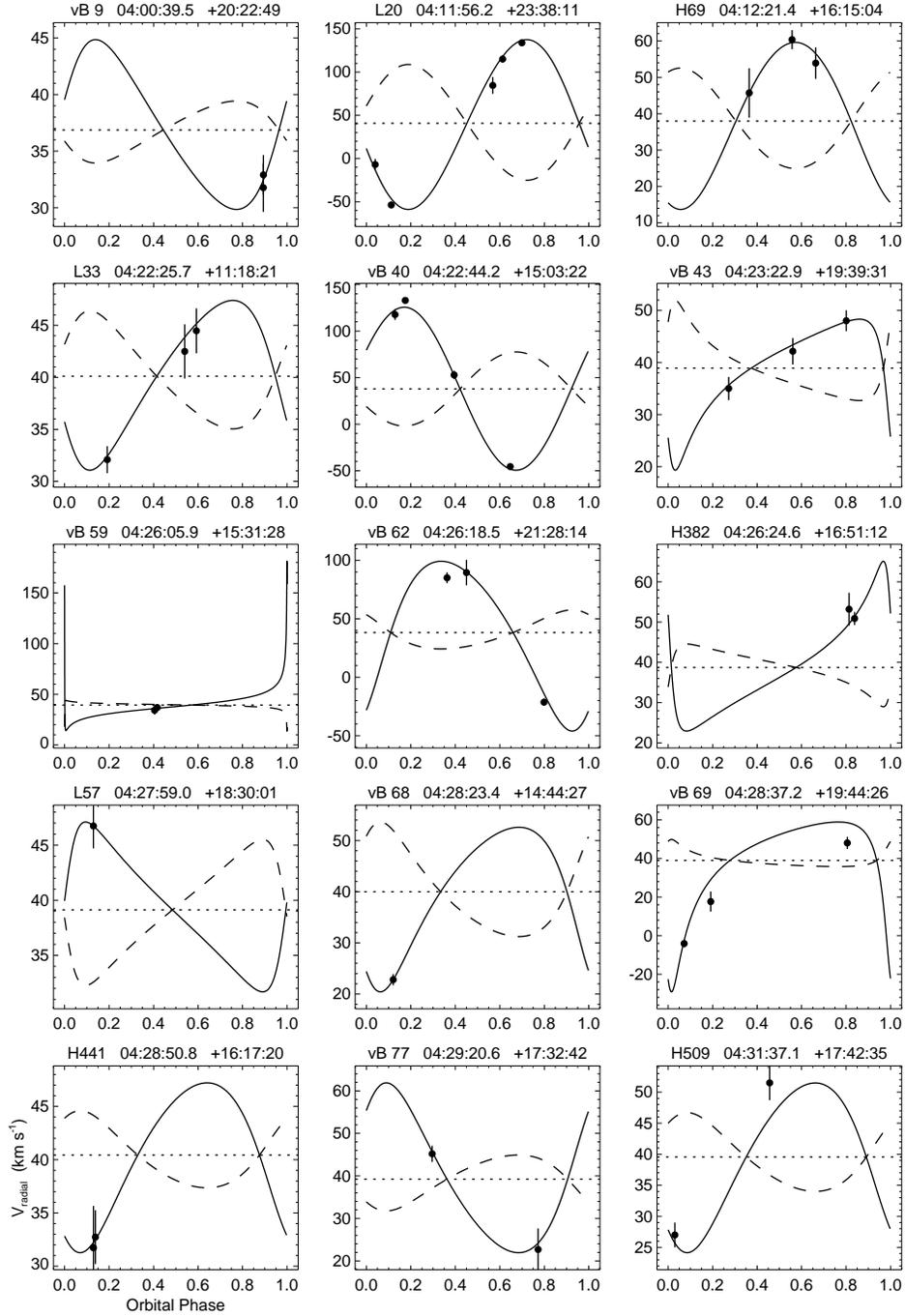}
  \caption{Double-lined velocity versus phase curves for the Hyades
    infrared sample, from R.A. $4^h00^m$ to $4^h31^m$.  The circles
    indicate the secondary velocities and associated uncertainties
    measured in the infrared.  For each binary, the dashed curve shows
    the CfA SB1 solution; the solid curve shows the SB2 solution.}
  \label{fig:sb2plots1}
\end{figure}

\begin{figure}
  \epsscale{0.8}
  \plotone{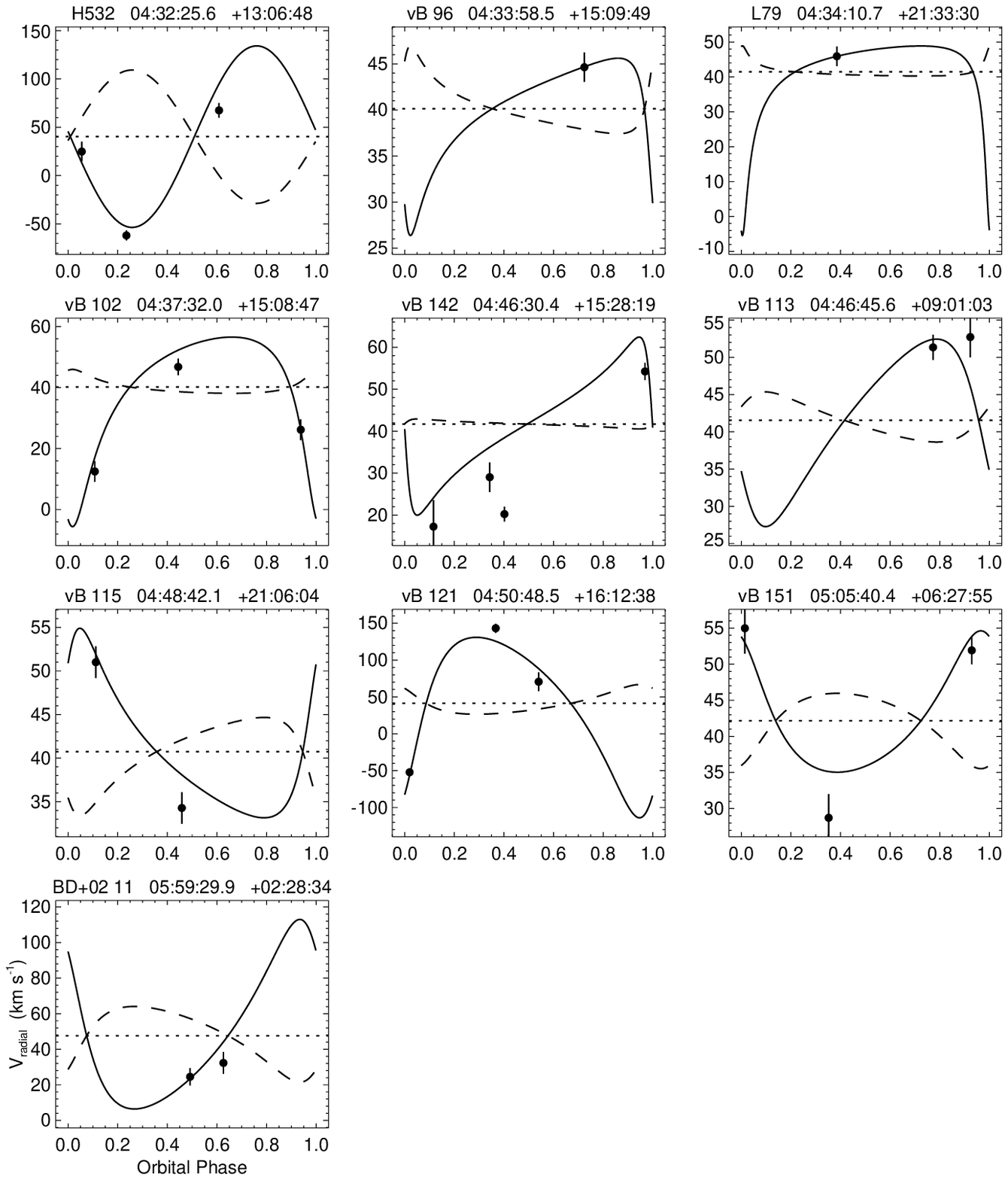}
  \caption{Same as Figure~\ref{fig:sb2plots1}, from R.A. $4^h32^m$ to
    $5^h59^m$.}
  \label{fig:sb2plots2}
\end{figure}

\begin{figure}
  \epsscale{0.8}
  \plotone{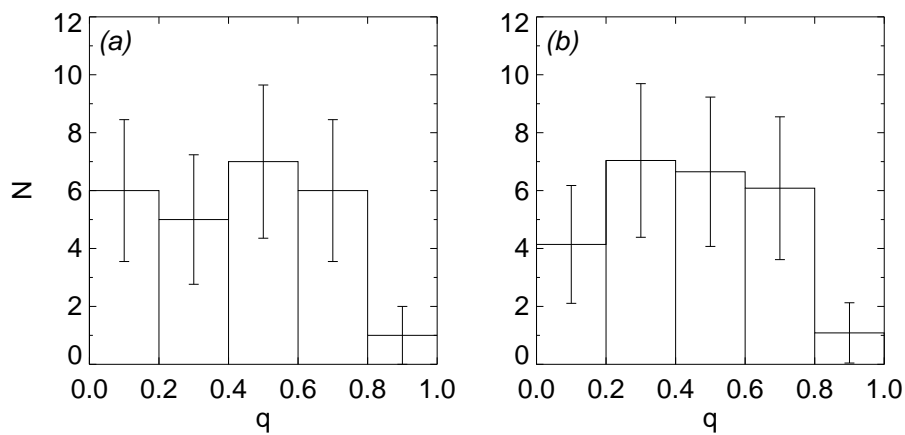}
  \caption{(\emph{a}) The distribution of mass ratios measured in the
    infrared for the SB2s in Table~\ref{table:orbitparams}. (\emph{b})
    Same as (\emph{a}), but includes the large variation in the
    uncertainties of our measured mass ratios. We distributed each
    mass ratio over a Gaussian with width equal to the $1\sigma$
    measurement uncertainty, clipped at
    $\q=\qmin-2\times\sigma_{\qmin}$ and $\q=1$, and normalized to
    unity.  The uncertainties shown for both (\emph{a}) and (\emph{b})
    are $\sqrt{N}$.}
  \label{fig:qhist}
\end{figure}

Figures~\ref{fig:sb2plots1} and \ref{fig:sb2plots2} show the SB2
velocity curves and measured secondary velocities, plotted against
orbital phase for the binaries in Table~\ref{table:orbitparams}.  The
primary velocity curves shown come directly from the SB1 solutions,
while the only free parameter in the secondary velocity curve is
\Ktwo. We show the secondary velocity uncertainties listed in
Table~\ref{table:obslog}, without the scaling correction described
earlier, because these values represent our best understanding of the
measurements.  As a consequence, some of the velocities are shown with
underestimated uncertainties.  Figure~\ref{fig:qhist} shows two
different representations of the mass ratio distribution for these
binaries: (\emph{a}) shows the case where each \q{} is determined with
equal precision; (\emph{b}) distributes each measured \q{} over its
corresponding uncertainty.  The similarity of the distributions,
within the reported $\sqrt{N}$ uncertainties, suggests that our
underestimate of the \Vtwo{} uncertainties has a small effect on the
overall distribution.  These distributions are obviously not complete
for the cluster, most importantly because it does not include the
large sample of binaries with $\q > 0.6$ that are detected as SB2s in
the CfA visible light spectroscopy.  Figure~\ref{fig:qhist} does,
however, clearly demonstrate the applicability of infrared
observations to binaries with small mass-ratios.

\subsection{\sc Alternate SB1 Parameters}
Alternate SB1 orbits are available in the literature for nine of our
infrared binaries: eight by R.\ F.\ Griffin and colleagues
\citep{griffin1978,griffin1981,griffin1985} and one by
\citet{sanford1921}.  The parameters published by these authors
provide additional verification of our assertion that the
uncertainties we report for \Ktwo{} and \q{} are dominated by our
infrared measurements of \Vtwo.  For each of these nine systems we
used the alternate SB1 parameters without modification to derive
\Ktwo{} and \q{}, following exactly the procedure describe above for
the CfA parameters.  However, we did not attempt to reconcile the
various velocity zero points, which have small differences of order
$1\kps$ or less \citep[e.g.,][]{detweiler1984}.
Table~\ref{table:altorbits} lists these SB1 parameters, their
references, and our alternate SB2 results.

The binaries L20, vB~43, vB~62, vB~69, vB~77, H509, and vB~121 have
\q{} listed in Tables~\ref{table:orbitparams} and
\ref{table:altorbits} that agree to $\sim1\sigma$ or better. The
parameters \Tknot{} and \w{} given for vB~40 by \citet{sanford1921}
have a $180^{\circ}$ phase difference from the CfA parameters, which
does not affect the derivation of \q.  The slightly greater than
$1\sigma$ difference in the derived \q{} for this binary could result
from differences in the velocity zero point; alternatively, vB~40
is a triple, and orbital motion of the wide pair over the
$\sim90$\,years between Sanford's measurement and our own could
account for the discrepancy.  The \q{} calculated for L57 using the
Griffin and CfA SB1 parameters demonstrate the importance of a
consistent velocity reference frame when combining SB1 and SB2
observations (\S\ref{irccorr}).  Due to its long period, L57 has a
small \Kone, which amplifies the effect of any offset in the velocity
reference frames.  Solving this system using the SB1 parameters of
\citet{griffin1985}, but with the CfA \g{}, yields $\q=0.85\pm0.24$,
nearly identical to the result in Table~\ref{table:orbitparams}.  This
issue is independent of the large uncertainties we report on \q{} for
L57, which result from our observing it only once in the infrared.

The results shown in Table~\ref{table:altorbits} confirm that the
uncertainties in our reported SB2 parameters originate primarily from
our infrared measurements, not from the underlying SB1 orbits.  The
unpublished CfA orbits provided a sample of SB1s that is
approximately three times larger than that available in the
literature, and that has a velocity reference frame which is not only
self-consistent, but also consistent with our infrared frame.

\section{\sc Discussion\label{discussion}}

\subsection{\sc Validation of SB2 Solutions\label{sb2validation}}

\begin{figure}
  \epsscale{1.0}
  \plotone{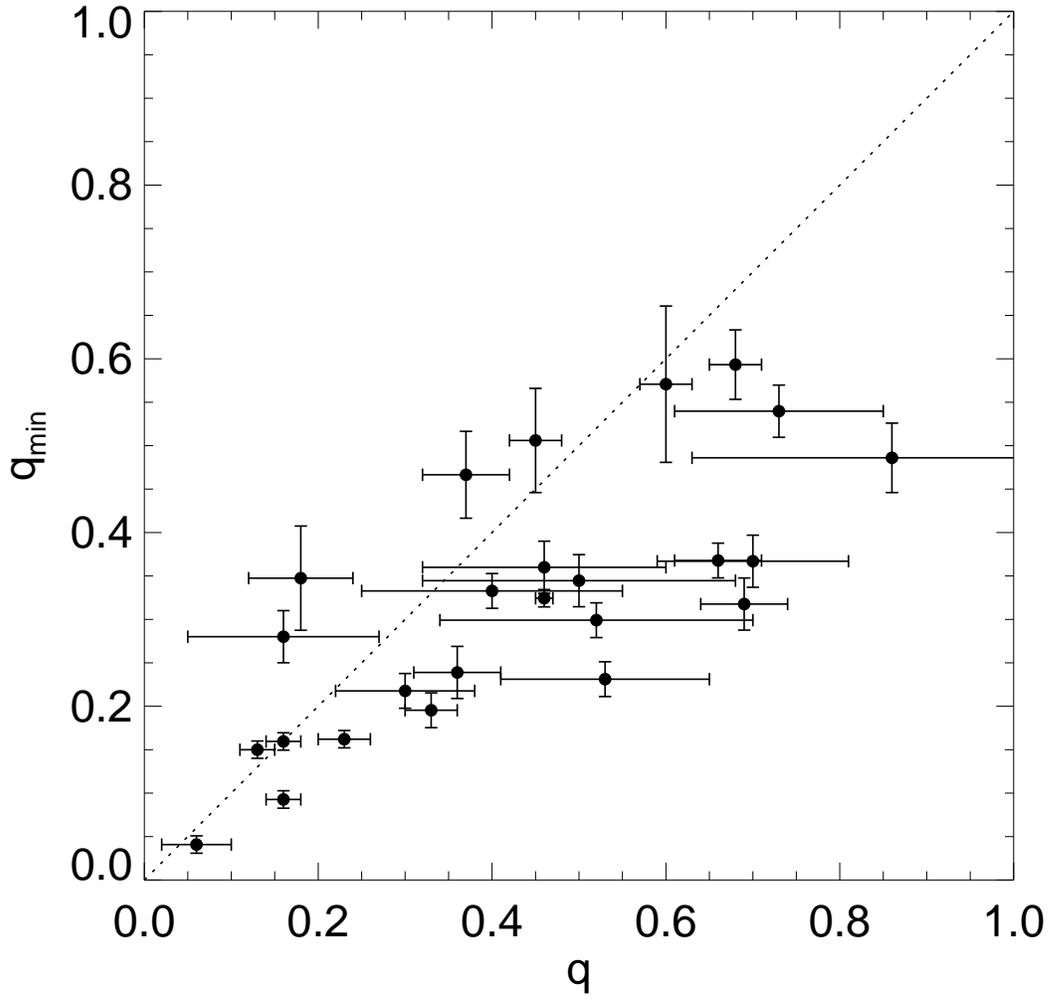}
  \caption{The minimum mass ratio, determined from \fmass, plotted
    against the measured mass ratio.  The dotted line indicates $\qmin
    = \q$.}
\label{fig:qvsqmin}
\end{figure}

Four of the SB2 solutions shown in Figures~\ref{fig:sb2plots1} and
\ref{fig:sb2plots2} depend on a single measurement of the secondary
velocity, and another nine on only two measurements.  Consequently,
many of the derived \Ktwo's, and the resulting \q's, have large
uncertainties.  Figure~\ref{fig:qvsqmin} plots \qmin{} against the
measured \q's to evaluate the plausibility of the measured values.  As
expected, \q{} is greater than or within $1\sigma$ of \qmin{} for all
of the binaries, except vB~59 and H382 which are consistent with \qmin
to better than $2\sigma$.  Both vB~59 and H382 have long periods and
were observed at orbital phases where the velocity separation was
small, resulting in low precision measurements of \Ktwo.  However, we
have two observations of each system at slightly different phases and
the measured secondary velocities are consistent.  Therefore, we are
confident that we are detecting the secondary in both of these
systems.  Observations of vB~59 in $\sim2012$ and of H382 as soon as
$\sim2009$, when the orbital phases will have changed significantly,
would further improve the SB2 solutions for these systems.

\begin{figure}
  \epsscale{1.0}
  \plotone{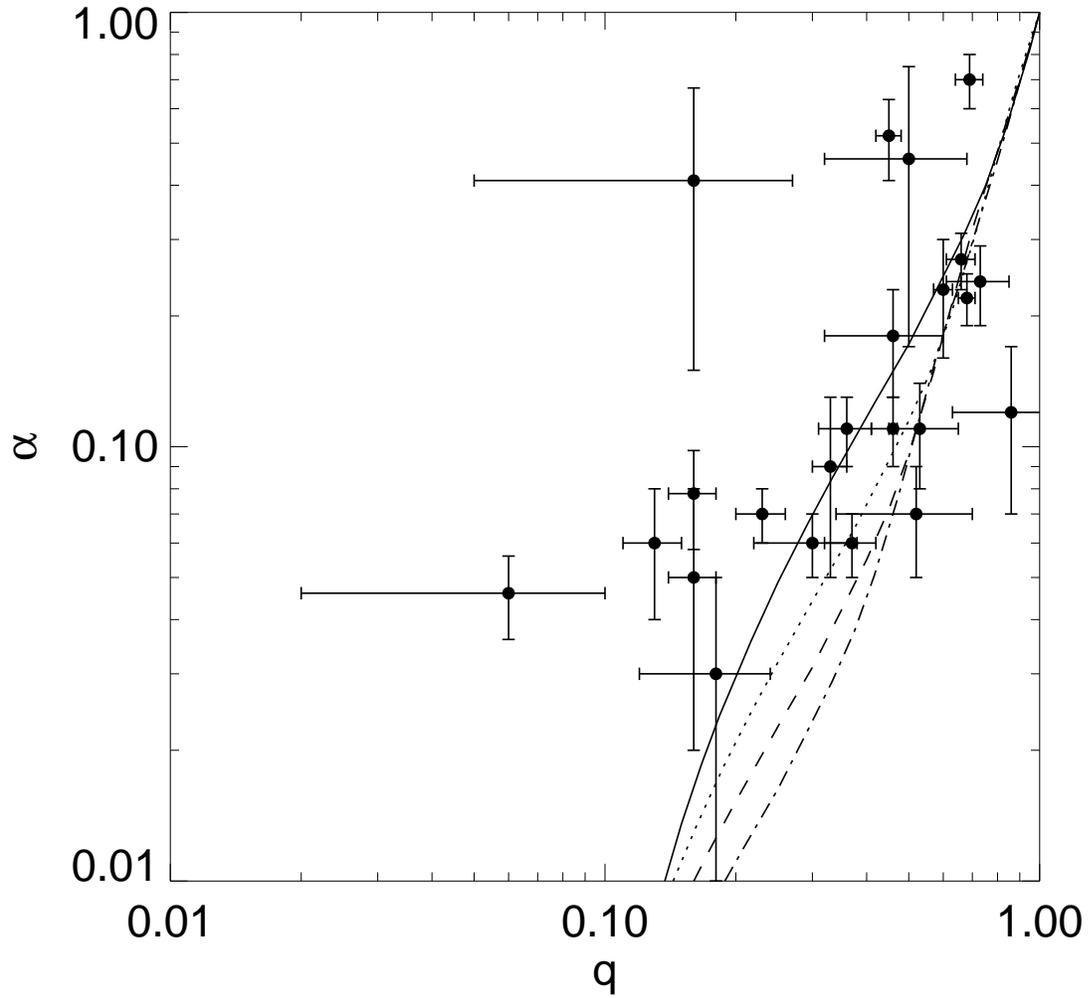}
  \caption{The measured flux ratio plotted against the measured mass
    ratio.  The curves show the theoretical \emph{H}-band flux ratios
    from \citet{baraffe1998} for 625 Myr old binaries with primary
    masses of 0.6\msun{} (\emph{solid}), 0.8\msun{} (\emph{dotted}),
    1.0\msun{} (\emph{dashed}), and 1.2\msun (\emph{dot-dashed}).}
  \label{fig:qvsalpha}
\end{figure}

The flux ratio, $\alpha$, measured by our cross-correlation routine
offers an additional validation of each SB2 solution.
Figure~\ref{fig:qvsalpha} plots the $\alpha$ measured for each binary,
averaged over all observations, against the measured mass ratio.  Also
shown are theoretical \emph{H}-band curves calculated from BCAH for
binaries with primary masses from 0.6\msun{} to 1.2\msun.  Two factors
complicate a direct comparison between our measured flux ratios and
the theoretical values.  First, the small wavelength range of CSHELL
spectra severely limits our ability to measure precise flux ratios
because individual spectral lines vary only a small amount relative to
each other with changing spectral type.  Considerably more accurate
flux ratios can be measured with spectra covering a larger wavelength
range, and thereby having many more spectral lines with differing
dependencies on spectral type.  Second, the theoretical curves
represent the integrated flux over the entire \emph{H}-band, of which
our CSHELL spectra only sample $\sim1.5\%$.  Nonetheless, our measured
values are well grouped along the curves.  The two obvious outliers
are L79 and vB~142.  We have only a single, low S/N observation of the
long period binary L79.  Our measured flux ratio is poorly
constrained, varying significantly depending on the exact pair of
templates used; the uncertainty shown in the figure probably
underestimates the actual uncertainty.  However, our measured
secondary velocity is well constrained, independent of the template
pair, and we are confident that the secondary detection is real.
vB~142 has a very small mass ratio, making the measurement of the
companion particularly difficult; \S\ref{vb142} addresses this system
in detail.

\subsection{\sc Infrared Non-Detections\label{sb2nondetections}}

\begin{figure}
  \epsscale{1.0}
  \plotone{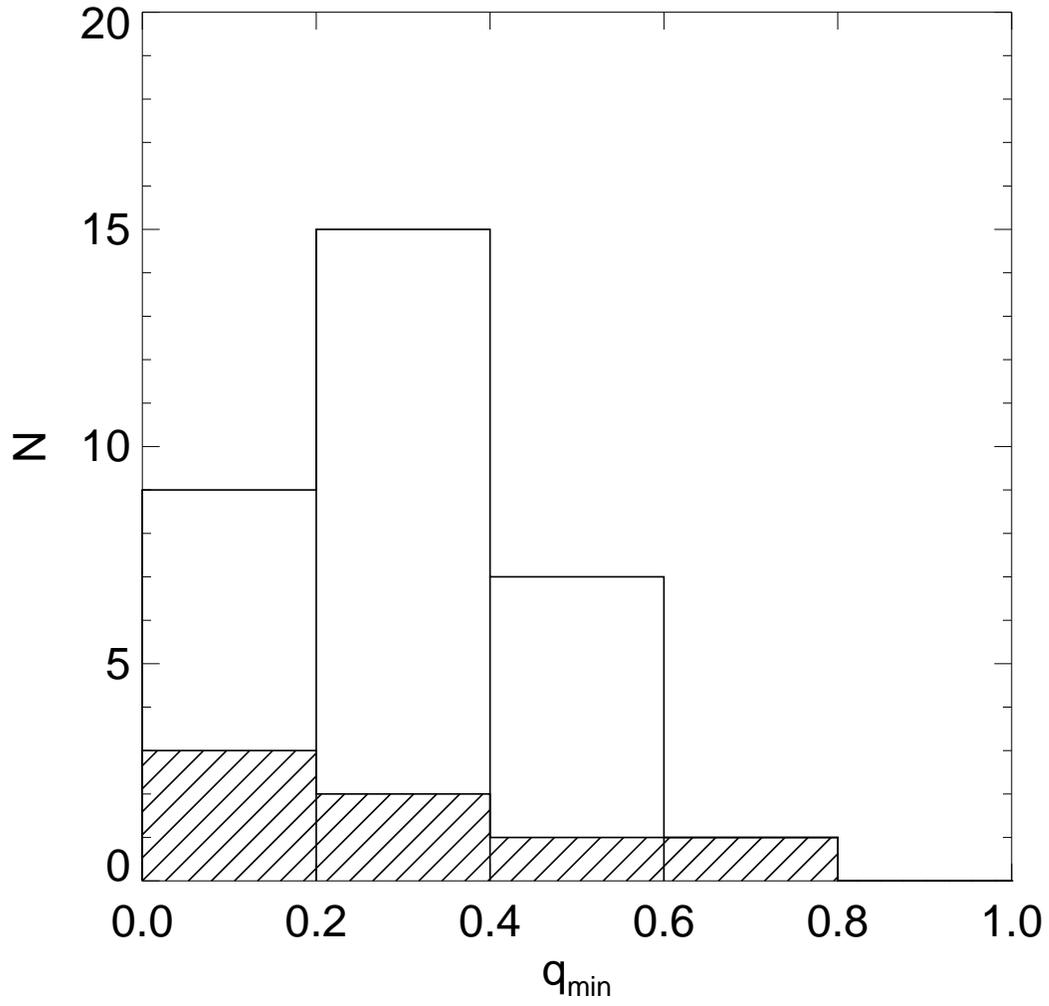}
  \caption{The distribution of \qmin{} for the systems detected
    (\emph{open}) and not detected (\emph{hashed}) as SB2s in the
    infrared.}
  \label{fig:qminhist}
\end{figure}

Our infrared observations failed to detect the secondary in seven
systems: vB~8, vB~30, vB~39, H411, L77, L90, and vB~114.
Figure~\ref{fig:qminhist} shows the distribution of \qmin{} for the
infrared sample binaries; the systems not detected as SB2s are
indicated by the hashed region.  We expected that our sensitivity to
binary companions would be incomplete for systems with the smallest
mass ratios because these systems also have small flux ratios.
However, several of the systems for which we did not detect the
secondary have large \qmin{}.  For three additional systems, vB~43,
L57, and H509, we did not detect the secondary in a subset of our
observations.  We address each of these ten systems below.

vB~8 and vB~30 have rapidly rotating F-type primary stars and both
have small \qmin{}, 0.15 and 0.19, respectively.  We observed both
spectra in the infrared at multiple epochs, with orbital phases where
the predicted velocity separation was large.  We propose two possible
explanations for our failure to detected these companions.  The true
value of \q{} may actually be close to \qmin.  For a given mass ratio,
the \emph{H}-band flux ratio of a binary decreases as the mass of the
primary increases.  Because vB~8 and vB~30 have primaries more massive
than those of a typical binary in our sample, their flux ratio's may
be too small to detect the secondaries at the S/N of our spectra.
Alternatively, if the companions are also rotating rapidly, their
spectral lines would be too broad to measure with CSHELL's limited
wavelength coverage, even with high S/N observations.

vB~114 and vB~39 have orbital periods of 4578 days and 5083 days,
respectively.  Our observations of both systems occurred at orbital
phases such that the primary velocities were near \g, and the
predicted velocity separations were only a few \kps.  Velocity
measurements under such conditions are inherently difficult, and even
had we detected the secondary components, the resulting \Ktwo's would
be poorly constrained.  vB~114 will have a more favorable orbital
phase in $\sim$2010; the phase of vB~39 will not improve until
$\sim$2013.  Both have moderate \qmin{}, 0.44 for vB~39 and 0.29 for
vB~114, so the flux ratios should not impede in detecting these
secondaries.

H411, L77, and L90 fall towards the faint edge of the distribution
shown in Fig.~\ref{fig:hmag}, and our observations of these targets
have S/N insufficient to detect their secondaries.  H411 has a small
\qmin, $\sim0.18$, which corresponds to a minimum H-band flux ratio of
only a few percent, and is at the limit of our best observation of
this binary, with S/N$\sim$50.  Because H411 is faint, obtaining a
spectrum with CSHELL that has better S/N would require many hours of
integration; such an observation could be carried out with a high
resolution spectrometer at an $8-10$ m telescope in a relatively small
amount of time.  L77 has a large \qmin{} of $\sim0.61$ and we expected
to be sensitive to its secondary.  However, our single observation of
L77 has S/N$\sim$25, and occurred at an orbital phase when the
velocity separation was small.  The orbit of this system is such that
even had we detected the secondary, the resulting SB2 orbit would be
largely unconstrained.  L77 will be at a more favorable orbital phase
in 2009.  L90 has $\qmin\sim0.32$; we observed it on two occasions,
each with S/N$\sim$50.  However, its long period and current orbital
phase indicate a small velocity separation.  This system will not be
at a more favorable orbital phase until $\sim$2011.

We did not detect a companion in our observations of vB~43 on 2005 Sep
30, L57 on 2005 Nov 25 and 2006 Feb 2, and H509 on 2005 Oct 2.  Each
of these observations were carried out at orbital phases corresponding
to primary velocities near \g. The observation of vB~43 was
compromised further by a low S/N, relative to our other observations
of it.  L57 has small \Kone{} and \Ktwo{}, which when combined with
it's K2 primary spectral type make it a difficult target to observe
with CSHELL: future observations would benefit from higher spectral
resolution or a larger free spectral range.

\subsection{\sc vB~142\label{vb142}}

Our four infrared observations of vB~142 provided good phase coverage,
and included measurements near the maximum velocity separation and on
both sides of the \g-velocity.  Despite this, the SB2 solution that we
derive is poorly determined and the residuals,
$V_2(fit)-V_2(measured)$, are large. We do not fully understand the
reasons behind this poor solution, but we include it here for
the following reasons.

Our observations, except for that on 2005 November 28, return
plausible velocities for the companion, albeit with large
uncertainties.  The 2005 November 28 observation occurred at phase
$\sim0.40$, and with small velocity separation, so our inability to
accurately measure the secondary in this spectrum is not surprising.
The average $\alpha$ that we measure for all of the observations,
$\sim0.04$, is, however, much larger than that predicted by BCAH
(Figure~\ref{fig:qvsalpha}).  To investigate this discrepancy we used
our M-type template LHS2351 to introduce an additional spectrum into
our observed vB~142 spectra from 2004 October 1 and 2005 February 22.
We added in this component with a ``true'' flux ratio,
$\alpha_{true}$, ranging from 0.05 to 0.0001, and, to avoid confusion
with the actual vB~142 companion, at a radial velocity of $-20\kps$. We
then attempted to recover this signal with our correlation procedure
and the set of templates used in the original vB~142 analyses,
excluding LHS2351.  We recovered the LHS2351 spectrum at the proper
velocity with $\alpha_{true}$ as small as 0.0005.  However, the
uncertainty of the measured velocity increased as $\alpha_{true}$
decreased, and the measured flux ratio, $\alpha_{meas.}$ became
unreliable for $\alpha_{true}\la0.01$.  Consider, for example, the
case of LHS2351 introduced into the 2005 February 22 spectrum with a
radial velocity of -20\kps{} and $\alpha_{true}=0.005$.  Our
correlation routine recovered this signal with a velocity of
$-25.1\pm7.0\kps$ and $\alpha_{meas.}=0.019\pm0.005$.  The larger than
expected $\alpha_{meas.}$ is consistent with the results obtained for
the vB~142 companion, and may arise because our primary templates do
not precisely match the vB~142 primary spectrum.  For small flux
ratios, the correlation routine tries to correct for this mismatch by
scaling the primary using $\alpha$; we have previously reported this
behavior \citep{bender2005}.

Our modeling with LHS2351 gives us confidence that we can detect a
companion with a very small flux ratio.  The \q{} that we derive for
vB~142, $0.06\pm0.04$, is consistent with $\qmin\sim0.04$ from the SB1
solution.  Whatever the true value of \q{} may be, our $3\sigma$ upper
limit of $\q\la0.18$ is very small.  Our primary goal in this endeavor
is to measure the binary mass ratio distribution for the Hyades (\S1):
\q{} for vB~142 is sufficiently well determined for this purpose.  Of
additional interest, the primary of vB~142 is a G5 star, and so a
companion with $\q=0.06\pm0.04$ could be a brown dwarf, which would be
an important discovery in the Hyades \citep{guenther2005}.

\subsection{\sc Component Masses\label{componentmasses}}

While spectroscopic observations of an SB2 yield its dynamical mass
ratio, they do not measure its orbital inclination and so alone they
cannot provide a dynamical measurement of the individual component
masses.  Observations of the visual orbit measure the inclination and
the total mass, and when combined with the mass ratio result in the
individual masses. All of the binaries in our sample have components
with a small angular separation and their visual orbits are not
currently available.  Those with periods longer than a few hundred
days are resolvable with adaptive optics imaging at a large aperture
telescope.

In the absence of visual orbits, we can still obtain good estimates of
the individual masses if the distance is known.  \emph{Hipparcos}
measured the parallax of most of our sample binaries
\citep{perryman1998}, and all have precise photometric measurements of
their total flux from 2MASS at J, H, and K.  We combined these with
our measured mass ratios and a theoretical mass-luminosity isochrone
from BCAH to calculate the individual component masses.  We chose the
BCAH models for several reasons: they show a good, albeit not perfect,
agreement with measured dynamical masses \citep{hillenbrand2004}; they
include the effects of atmospheres, which are important in the low
mass regime that applies to most of our secondaries; and lastly, they
are provided in a convenient form that specifies magnitudes in the
standard photometric bands used by observers.  We used the 625 Myr
isochrone, while noting that at such an old age the mass-luminosity
relationship is mostly insensitive to age.
Table~\ref{table:componentmasses} lists the calculated component
masses.

The uncertainties given in Table~\ref{table:componentmasses} include
contributions from the parallax, photometry, and mass ratio; they do
not include any uncertainties from the BCAH models.  All of the 2MASS
J, H, and K photometric uncertainties are small, $\sim0.02-0.03$ mag.
Because our binaries have small flux ratios, the precision with which
we determine the primary masses is mostly dependent on the precision
of the parallax measurements.  The uncertainty for the secondaries
strongly depends on the precision of the mass ratios.  Most of the
primaries have masses determined to better than 10\%, while some of
the secondaries approach this level.  \emph{Hipparcos} did not measure
the parallax of vB~59 or L57, so for these systems we used values
reported in the Tycho catalog, and the resulting uncertainties on both
the primary and secondary masses are large.  The parallaxes of H69,
H441, and L79 have not been measured, so we estimated their primary
masses directly from their spectral type and assumed an uncertainty of
$\sim0.1\msun$.  The secondary masses then follow directly from our
measured mass ratios.  Finally, the primary of vB~68 is more massive
than the range covered by BCAH, so for this system only we used the
empirical isochrone determined by \citet{pinsonneault2004} and note
that the resulting masses are consistent with measured spectral types.

\section{\sc Summary\label{summary}} 

We have obtained high resolution infrared spectroscopy of 32 SBs in
the Hyades, whose SB1 orbital parameters have been measured by RPS and
DWL at the CfA, in order to detect their companions and thereby study
the binary mass ratio distribution in this young cluster.  We detected
the companion in 25 of these systems.  For these, we combined our
results with the SB1 parameters to determine their solutions as SB2s.
Some of the SB2 solutions we report have low precision for \Ktwo{}.
However, obtaining precise orbital parameters for individual systems
was not our objective here.  Instead, our intent was to constrain the
\emph{distribution} of mass ratios in binaries with low mass companions, and
our results are sufficient for this purpose.  We also estimated the
primary and secondary masses of our sample binaries using 2MASS
photometry, \emph{Hipparcos} parallax measurements, and our measured
mass ratios.  The mass ratios of the binaries with the most reliable
SB2 solutions span the range from $\q\sim0.1-0.8$, corresponding to
secondary masses as small as $\sim0.15\msun$.  We also detect a very
low mass companion to vB~142.  The solution for its mass ratio is not
yet reliable, but it appears to be $\q\la0.18$ at the $3\sigma$ level,
and may represent the detection of a brown dwarf companion.

The precision of our derived primary masses is limited by the
uncertainties in the \emph{Hipparcos} parallax measurements.  The
secondary mass measurements, however, can be improved significantly by
reducing the uncertainties on the measured mass ratios through
additional infrared observations.  Direct observations of sample
members as visual binaries would measure their orbital inclinations
and total masses, and when combined with the spectroscopy would yield
dynamical component masses.  By utilizing the \emph{Hipparcos}
distances, such measurements would contribute a test of the
theoretical mass-luminosity relationships \citep[e.g.,][]{mathieu2007}.
Improving the SB2 solutions or measuring the visual orbits of our
sample would require a significant commitment of observing time and
analysis resources.  The visual orbit mapping may require
technological improvements in interferometry and adaptive optics
techniques.

The results presented here are an initial step in a program to produce
distributions of mass-ratio and secondary mass for Hyades cluster
binaries with periods from a few days to a few thousand days. Future
papers in this series will combine our new determinations of mass
ratios with available orbital solutions from other Hyades binary
surveys, including the the spatially resolved systems studied by
\citet{patience1998}, to present the mass ratio and secondary mass
distributions for the cluster.  We also intend to address more fully the
set of hierarchical triple systems, for which, when combined with the
speckle observations of \citet{patience1998}, we have information on
both the inner and outer orbits.

\acknowledgments 

We are grateful to DWL and RPS for providing the CfA SB1 parameters
and for numerous discussions that improved the manuscript.  We also
thank the referee for several educational suggestions concerning the
secondary velocity precision.  We thank L. Prato for providing the
NIRSPEC observations, T. Mazeh for suggesting the procedure used to
estimate the velocity uncertainties, and the telescope operators and staff
at the IRTF for their support during our many observing runs. The
authors are visiting astronomers at the Infrared Telescope Facility,
which is operated by the University of Hawaii under Cooperative
Agreement no.\ NCC 5-538 with the National Aeronautics and Space
Administration, Science Mission Directorate, Planetary Astronomy
Program.  CB is supported by an NRC Research Associateship Award at
NRL.  Basic research in infrared astronomy at NRL is supported by 6.1
base funding.  The authors were supported at Stony Brook in part by
NSF grants 02-05427 and 06-07612.  Data presented herein were obtained
at the W.M.\ Keck Observatory, which is operated as a scientific
partnership among the California Institute of Technology, the
University of California and the National Aeronautics and Space
Administration. The Observatory was made possible by the generous
financial support of the W.M.\ Keck Foundation.  This research made
use of the SIMBAD database, operated at CDS, Strasbourg, France, data
products from 2MASS, which is a joint project of the University of
Massachusetts and IPAC at the California Institute of Technology,
funded by NASA and NSF, and the Hipparcos and Tycho Catalogues, ESA
SP-1200.  The authors wish to extend special thanks to those of
Hawaiian ancestry on whose sacred mountain we are privileged to be
guests.

{\it Facilities:} \facility{IRTF ()},\facility{Keck:II ()}

%tables

\clearpage
\begin{deluxetable}{lcccccc}
\tablecaption{Hyades Infrared Sample\label{table:irsample}}
\tablewidth{0pt}
\tablecolumns{7}
\tabletypesize{\footnotesize}
\tablehead{
%  \colhead{} & \colhead{} & \colhead{R.A.} & \colhead{Dec.} &
%  \colhead{} & \colhead{} & \colhead{} \\%& 
  %\multicolumn{2}{c}{$\mathrm{N_{obs}}$}\\ 
  \colhead{Target} & \colhead{HD} & \colhead{R.A. (J2000)} &
  \colhead{Dec. (J2000)} & \colhead{Sp. Type} & \colhead{\emph{V}} &
  \colhead{\emph{H}} }%& \colhead{Visible} & \colhead{IR} }
\startdata
vB 8       & \phn25102   &  03 59 40.49 & +10 19 49.4 & F5 & \phn6.4  & 5.4 \\
vB 9       & \nodata     &  04 00 39.54 & +20 22 49.5 & G4 & \phn8.7  & 7.0 \\
L20        & 284163      &  04 11 56.22 & +23 38 10.8 & K0 & \phn9.4  & 6.6 \\
H69        & \nodata     &  04 12 21.44 & +16 15 03.5 & M1 & 14.0     & 9.1 \\
vB 30      & \phn27397   &  04 19 57.70 & +14 02 06.7 & F0 & \phn5.6  & 4.9 \\
L33        & 286770      &  04 22 25.69 & +11 18 20.6 & K8 & \phn9.8  & 7.1 \\
vB 40      & \phn27691   &  04 22 44.17 & +15 03 21.9 & G0 & \phn7.0  & 5.6 \\
vB 39      & \phn27685   &  04 22 44.78 & +16 47 27.7 & G4 & \phn7.8  & 6.3 \\
vB 43      & 284414      &  04 23 22.85 & +19 39 31.2 & K2 & \phn9.4  & 7.3 \\
vB 59      & \phn28034   &  04 26 05.86 & +15 31 27.6 & G8 & \phn7.5  & 6.2 \\
vB 62      & \phn28033   &  04 26 18.50 & +21 28 13.6 & F8 & \phn7.4  & 6.1 \\
H382       & \phn28068   &  04 26 24.61 & +16 51 12.0 & G1 & \phn8.0  & 6.6 \\
H411       & 285828      &  04 27 25.34 & +14 15 38.5 & K2 & 10.3     & 7.8 \\
L57        & 285766      &  04 27 58.96 & +18 30 00.9 & K2 & 10.2     & 7.7 \\
vB 68      & \phn28294   &  04 28 23.40 & +14 44 27.5 & F0 & \phn5.9  & 5.1 \\
vB 69      & \phn28291   &  04 28 37.21 & +19 44 26.5 & G5 & \phn8.6  & 7.0 \\
H441       & 285806      &  04 28 50.81 & +16 17 20.3 & K7 & 10.7     & 7.6 \\
vB 77      & \phn28394   &  04 29 20.55 & +17 32 41.8 & F7 & \phn7.0  & 5.8 \\
H509       & \phn28634   &  04 31 37.10 & +17 42 35.2 & K2 & \phn9.5  & 7.3 \\
H532       & 286839      &  04 32 25.65 & +13 06 47.6 & K0 & 11.0     & 7.8 \\
vB 96      & 285931      &  04 33 58.54 & +15 09 49.0 & K0 & \phn8.5  & 6.6 \\
L79        & \nodata     &  04 34 10.73 & +11 33 29.6 & K7 & 11.7     & 8.3 \\
L77        & \nodata     &  04 34 49.76 & +20 23 41.6 & K7 & 11.1     & 8.0 \\
vB 102     & \phn29310   &  04 37 31.98 & +15 08 47.2 & G1 & \phn7.5  & 6.1 \\
L90        & \phn29896   &  04 43 15.70 & +17 04 08.8 & K0 & \phn9.9  & 7.5 \\
vB 142     & \phn30246   &  04 46 30.39 & +15 28 19.4 & G5 & \phn8.3  & 6.8 \\
vB 113     & \phn30311   &  04 46 45.58 & +09 01 02.7 & F5 & \phn7.2  & 5.9 \\
vB 114     & \phn30355   &  04 47 37.57 & +18 15 31.4 & G0 & \phn8.5  & 6.9 \\
vB 115     & 284787      &  04 48 42.12 & +21 06 03.6 & G5 & \phn9.1  & 7.2 \\
vB 121     & \phn30738   &  04 50 48.54 & +16 12 37.6 & F8 & \phn7.3  & 6.2 \\
vB 151     & 240692      &  05 05 40.38 & +06 27 54.6 & K2 & \phn9.9  & 7.6 \\
BD+02 1102 & \phn40512   &  05 59 29.92 & +02 28 34.2 & F5 & \phn7.8  & 6.7 \\
\enddata
\end{deluxetable}     

\begin{deluxetable}{llccc}
  \tablecaption{Log of Infrared Observations\label{table:obslog}}
  \tablewidth{0pt}
  \tablecolumns{5}
  \tabletypesize{\footnotesize}
  \tablehead{
    \colhead{} & \colhead{} & \colhead{} & \colhead{} & \colhead{\Vtwo} \\
    \colhead{Target} & \colhead{UT Date} & \colhead{JD-2400000} & 
    \colhead{Instrument} & \colhead{(\kps)}}
  \startdata
vB 8              & 2004 Oct 4  & 53282.967 & C & \nodata               \\
                  & 2005 Sep 30 & 53643.982 & C & \nodata               \\
vB 9              & 2004 Oct 2  & 53280.978 & C & $\phnp31.8\pm\phn2.1$ \\
                  & 2004 Oct 4  & 53282.993 & C & $\phnp32.9\pm\phn1.8$ \\ 
L20               & 2004 Oct 1  & 53279.959 & C & $\php133.9\pm\phn4.0$ \\
                  & 2004 Oct 2  & 53280.950 & C & $\phn-53.6\pm\phn3.6$ \\
                  & 2004 Oct 3  & 53282.148 & C & $\php115.6\pm\phn4.3$ \\
                  & 2005 Oct 10 & 53645.985 & C & $\phnp84.6\pm\phn9.7$ \\
                  & 2005 Oct 3  & 53647.112 & C & $\phnn-6.9\pm\phn6.5$ \\
H69               & 2004 Oct 3  & 53282.999 & C & $\phnp60.4\pm\phn2.6$ \\ 
                  & 2005 Feb 22 & 53423.715 & N & $\phnp53.9\pm\phn4.3$ \\ 
                  & 2006 Feb 3  & 53769.734 & C & $\phnp45.7\pm\phn6.8$ \\
vB 30             & 2004 Oct 3  & 53282.035 & C & \nodata               \\
                  & 2004 Oct 4  & 53283.036 & C & \nodata               \\ 
L33               & 2004 Oct 3  & 53281.959 & C & $\phnp32.1\pm\phn1.3$ \\
                  & 2005 Oct 2  & 53646.018 & C & $\phnp42.5\pm\phn2.6$ \\
                  & 2005 Nov 26 & 53700.826 & C & $\phnp44.5\pm\phn2.2$ \\
vB 40             & 2004 Oct 1  & 53280.012 & C & $\phn-45.5\pm\phn1.9$ \\ 
                  & 2004 Oct 3  & 53282.123 & C & $\php132.9\pm\phn3.0$ \\ 
                  & 2005 Oct 2  & 53645.958 & C & $\php117.7\pm\phn5.8$ \\
                  & 2005 Oct 3  & 53647.020 & C & $\phnp52.9\pm\phn4.6$ \\ 
vB 39             & 2005 Nov 26 & 53700.897 & C & \nodata \\
                  & 2006 Feb 2  & 53768.727 & C & \nodata \\
vB 43             & 2004 Oct 3  & 53282.052 & C & $\phnp42.2\pm\phn2.5$ \\ 
                  & 2005 Feb 22 & 53423.754 & N & $\phnp48.0\pm\phn2.0$ \\ 
                  & 2005 Sep 30 & 53644.008 & C & \nodata \\ 
                  & 2005 Nov 27 & 53701.755 & C & $\phnp35.0\pm\phn2.2$ \\ 
vB 59             & 2005 Nov 27 & 53701.813 & C & $\phnp34.2\pm\phn4.4$ \\ 
                  & 2006 Feb 2  & 53768.812 & C & $\phnp36.9\pm\phn2.9$ \\
vB 62             & 2004 Oct 1  & 53280.052 & C & $\phnp85.1\pm\phn4.6$ \\
                  & 2005 Nov 25 & 53699.778 & C & $\phnp89.6\pm10.9$ \\
	          & 2005 Nov 28 & 53702.758 & C & $\phn-21.4\pm\phn3.2$ \\
H382              & 2005 Nov 28 & 53702.801 & C & $\phnp53.2\pm\phn4.0$ \\
                  & 2006 Feb 2  & 53768.746 & C & $\phnp50.9\pm\phn1.2$ \\
H411              & 2004 Oct 3  & 53282.078 & C & \nodata               \\
                  & 2005 Oct 2  & 53646.060 & C & \nodata               \\
L57               & 2004 Oct 3  & 53282.104 & C & $\phnp46.7\pm\phn2.0$ \\
                  & 2005 Nov 25 & 53699.828 & C & \nodata \\
	          & 2006 Feb 2  & 53768.889 & C & \nodata \\
vB 68\tablenotemark{a}& 2005 Oct 3  & 53657.074 & C & $\phnp22.8\pm\phn1.1$ \\
vB 69             & 2004 Oct 1  & 53280.082 & C & $\phnn-4.1\pm\phn2.0$ \\ 
                  & 2005 Sep 30 & 53644.039 & C & $\phnp48.1\pm\phn3.1$ \\
	          & 2005 Nov 27 & 53701.787 & C & $\phnp17.7\pm\phn5.2$ \\
H441              & 2005 Nov 28 & 53702.851 & C & $\phnp31.7\pm\phn3.9$ \\ 
                  & 2006 Feb 2  & 53768.845 & C & $\phnp32.7\pm\phn2.6$ \\
vB 77\tablenotemark{a}& 2004 Oct 1  & 53280.010 & C & $\phnp22.7\pm\phn4.9$ \\
                  & 2005 Nov 30 & 53644.122 & C & $\phnp45.2\pm\phn1.9$ \\
H509              & 2004 Oct 2  & 53281.117 & C & $\phnp51.5\pm\phn2.8$ \\
                  & 2005 Oct 2  & 53646.099 & C & \nodata \\
                  & 2006 Feb 2  & 53768.786 & C & $\phnp27.0\pm\phn2.0$ \\
H532              & 2004 Oct 2  & 53281.015 & C & $\phn-61.9\pm\phn5.3$ \\
                  & 2004 Oct 4  & 53283.055 & C & $\phnp67.5\pm\phn7.5$ \\
                  & 2005 Nov 26 & 53700.916 & C & $\phnp24.9\pm10.2$ \\
vB 96             & 2006 Feb 3  & 53769.922 & C & $\phnp44.6\pm\phn1.6$ \\
L79               & 2006 Feb 3  & 53769.792 & C & $\phnp45.9\pm\phn2.8$ \\
L77               & 2006 Feb 3  & 53769.840 & C &  \nodata \\
vB 102            & 2004 Oct 2  & 53281.041 & C & $\phnp46.8\pm\phn2.7$ \\
                  & 2005 Sep 30 & 53644.068 & C & $\phnp26.2\pm\phn3.4$ \\
                  & 2006 Feb 2  & 53768.921 & C & $\phnp12.5\pm\phn3.4$ \\
L90               & 2005 Nov 25 & 53699.881 & C & \nodata               \\
                  & 2006 Feb 3  & 53769.889 & C & \nodata               \\
vB 142            & 2004 Oct 1  & 53280.118 & C & $\phnp54.2\pm\phn2.1$ \\
                  & 2005 Feb 22 & 53423.860 & N & $\phnp17.2\pm\phn6.4$ \\
                  & 2005 Oct 1  & 53645.099 & C & $\phnp29.0\pm\phn3.5$ \\
                  & 2005 Nov 28 & 53702.910 & C & $\phnp20.2\pm\phn1.8$ \\
vB 113            & 2004 Oct 2  & 53281.066 & C & $\phnp50.7\pm\phn1.4$ \\
                  & 2005 Oct 1  & 53645.061 & C & $\phnp51.5\pm\phn3.4$ \\
vB 114            & 2005 Nov 27 & 53701.930 & C & \nodata               \\
vB 115            & 2004 Oct 4  & 53283.082 & C & $\phnp51.0\pm\phn1.8$ \\
                  & 2005 Nov 27 & 53701.894 & C & $\phnp34.3\pm\phn1.8$ \\
vB 121            & 2004 Oct 1  & 53280.135 & C & $\phn-52.2\pm\phn3.4$ \\ 
                  & 2004 Oct 3  & 53282.134 & C & $\php143.2\pm\phn6.2$ \\
                  & 2005 Nov 28 & 53702.941 & C & $\phnp70.6\pm12.7$ \\
vB 151            & 2004 Oct 4  & 53283.112 & C & $\phnp28.7\pm\phn3.3$ \\
                  & 2005 Oct 2  & 53646.135 & C & $\phnp51.9\pm\phn1.9$ \\
                  & 2005 Nov 25 & 53699.929 & C & $\phnp54.9\pm\phn3.5$ \\
BD+02 1102\tablenotemark{a}& 2004 Oct 4  & 53283.136 & C & $\phnp24.5\pm\phn4.8$ \\ 
                  & 2005 Oct 1  & 53645.135 & C & $\phnp32.3\pm\phn6.2$ \\
\enddata
\tablenotetext{a}{Indicates a rapidly rotating primary.  See
  \S\ref{irrapidrotators} for a discussion.}
\end{deluxetable}

\begin{deluxetable}{lcccccccccccc}
  \tablecaption{Orbital Solutions\label{table:orbitparams}}
  \tablewidth{0pt}
  \tablecolumns{13}
  \tabletypesize{\scriptsize}
  \setlength{\tabcolsep}{0.025in}
  \rotate
  \tablehead{
    \colhead{} & 
    \colhead{\period} & 
    \colhead{} &
    \colhead{\w} &
    \colhead{\Tknot} &
    \colhead{\g} & 
    \colhead{\Kone} & 
    \colhead{\Ktwo} &
    \colhead{} &
    \colhead{$a_1\sin i$} &
    \colhead{$a_2\sin i$} &
    \colhead{$M_1\sin^3i$} &
    \colhead{$M_2\sin^3i$} \\
     
    \colhead{Target} & \colhead{(days)} & \colhead{\e} &
    \colhead{(deg)} & \colhead{(JD-2400000)} & \colhead{(\kps)} &
    \colhead{(\kps)} & \colhead{(\kps)} & \colhead{\q}  & 
    \colhead{(Gm)} & 
    \colhead{(Gm)} & 
    \colhead{(\msun)} &
    \colhead{(\msun)}} 
    \startdata
  vB 9       & 5070\phdnnnnnn     & 0.22\phnn & 107\phdn & 48747\phdnnn & 36.87\phn & \phn2.74 & $\phnn7.50\pm\phn0.92$       & $0.36\pm0.05$ &  $186\phdnnnn\pm23\phdnnnn$       & $\phn510\phdnn\pm\phn84\phdnn$ & $0.38\phn\pm0.13\phn$ & $0.140\phn\pm0.034\phn$\\
  L20        & \phnnn2.394358     & 0.053\phn & 286.4    & 50553.507    & 40.86\phn & 66.96    & $\phn98.2\phn\pm\phn3.6\phn$ & $0.68\pm0.03$ &  $\phnn2.202\phn\pm\phn0.014\phn$ & $\phnnn3.23\pm\phnn0.12$       & $0.661\pm0.053$       & $0.451\phn\pm0.020\phn$\\
  H69        & \phn128.114\phnnn  & 0.062\phn & 337\phdn & 49110.9\phnn & 37.96\phn &  13.80   & $\phn22.9\phn\pm\phn1.2\phn$ & $0.60\pm0.03$ &  $\phn24.26\phnn\pm\phn0.56\phnn$ & $\phnn40.3\phn\pm\phnn2.1\phn$ & $0.407\pm0.048$       & $0.245\phn\pm0.019\phn$\\
  L33        & 1044.9\phnnnnn     & 0.250\phn & 295.5    & 48902\phdnnn & 40.11\phn & \phn5.67 & $\phnn8.16\pm\phn0.58$       & $0.69\pm0.05$ &  $\phn78.9\phnn\pm\phn2.6\phnnn$  & $\phn113.5\phn\pm\phnn8.1\phn$ & $0.153\pm0.024$       & $0.106\phn\pm0.011\phn$\\
  vB 40      & \phnnn4.000177     & 0.0043    & 118\phdn & 48353.24\phn & 37.920    & 39.88    & $\phn87.5\phn\pm\phn2.7\phn$ & $0.46\pm0.02$ &  $\phnn2.1934\pm\phn0.0042$       & $\phnnn4.81\pm\phnn0.15$       & $0.588\pm0.043$       & $0.268\phn\pm0.011\phn$\\
  vB 43      & \phn589.76\phnnnn  & 0.619\phn & 304.7    & 50002.65\phn & 38.925    & \phn9.61 & $\phn14.5\phn\pm\phn1.2\phn$ & $0.66\pm0.05$ &  $\phn61.23\phnn\pm\phn0.78\phnn$ & $\phnn92.4\phn\pm\phnn7.4\phn$ & $0.249\pm0.046$       & $0.165\phn\pm0.017\phn$\\
  vB 59      & 5724\phdnnnnnn     & 0.975\phn & 224.2    & 51385\phdnnn & 39.349    & 15.1\phn & $\phn84\phdnn\pm28\phdnn$    & $0.18\pm0.06$ &  $264\phdnnnn\pm28\phdnnnn$       & $1460\phdnn\pm490\phdnn$       & $5.4\phnn\pm4.8\phnn$ & $0.96\phnn\pm0.56\phnn$\\
  vB 62      & \phnnn8.550647     & 0.212\phn & \phn41.0 & 49925.248    & 38.247    & 16.73    & $\phn72.6\phn\pm\phn8.8\phn$ & $0.23\pm0.03$ &  $\phnn1.923\phn\pm\phn0.017\phn$ & $\phnnn8.3\phn\pm\phnn1.0\phn$ & $0.48\phn\pm0.15\phn$ & $0.110\phn\pm0.022\phn$\\
  H382       & 2657\phdnnnnnn     & 0.682\phn & 248.4    & 48884.7\phnn & 38.73\phn & \phn7.79 & $\phn21.1\phn\pm\phn2.4\phn$ & $0.37\pm0.05$ &  $208\phdnnnn\pm11\phdnnnn$       & $\phn563\phdnn\pm\phn66\phdnn$ & $1.90\phn\pm0.55\phn$ & $0.70\phnn\pm0.14\phnn$\\
  L57        & 1911.3\phnnnnn     & 0.486\phn & \phn94.1 & 49211.6\phnn & 39.124    & \phn6.63 & $\phnn7.7\phn\pm\phn2.1\phn$ & $0.86\pm0.23$ &  $152.2\phnnn\pm\phn2.6\phnnn$    & $\phn177\phdnn\pm\phn48\phdnn$ & $0.21\phn\pm0.21\phn$ & $0.180\phn\pm0.053\phn$\\
  vB 68      & \phn331.66\phnnnn  & 0.288\phn & 319\phdn & 50290.4\phnn & 40.00\phn & 11.19    & $\phn16.1\phn\pm\phn2.4\phn$ & $0.70\pm0.11$ &  $\phn48.9\phnnn\pm\phn2.5\phnnn$ & $\phnn70\phdnn\pm\phn11\phdnn$ & $0.36\phn\pm0.12\phn$ & $0.251\phn\pm0.051\phn$\\
  vB 69      & \phnn41.6729\phnn  & 0.643\phn & 328.4    & 49443.176    & 38.993    & \phn7.03 & $\phn44.0\phn\pm\phn5.9\phn$ & $0.16\pm0.02$ &  $\phnn3.08\phnn\pm\phn0.11\phnn$ & $\phnn19.3\phn\pm\phnn2.6\phn$ & $0.222\pm0.082$       & $0.0355\pm0.0085$      \\
  H441       & 7494\phdnnnnnn     & 0.186\phn & 324\phdn & 52730\phdnnn & 40.44\phn & \phn3.62 & $\phnn8.0\phn\pm\phn0.3\phn$ & $0.45\pm0.03$ &  $366\phdnnnn\pm17\phdnnnn$       & $\phn807\phdnn\pm\phn36\phdnn$ & $0.796\pm0.073$       & $0.360\phn\pm0.033\phn$\\
  vB 77      & \phn238.86\phnnnn  & 0.200\phn & 132\phdn & 48557.3\phnn & 39.22\phn & \phn6.58 & $\phn20.0\phn\pm\phn1.8\phn$ & $0.33\pm0.03$ &  $\phn21.18\phnn\pm\phn0.91\phnn$ & $\phnn64.3\phn\pm\phnn5.7\phn$ & $0.329\pm0.075$       & $0.108\phn\pm0.016\phn$\\
  H509       & \phn849.95\phnnnn  & 0.174\phn & 316.9    & 48643\phdnnn & 39.553    & \phn6.35 & $\phn13.7\phn\pm\phn4.0\phn$ & $0.46\pm0.14$ &  $\phn73.0\phnnn\pm\phn1.5\phnnn$ & $\phn157\phdnn\pm\phn46\phdnn$ & $0.46\phn\pm0.32\phn$ & $0.215\phn\pm0.086\phn$\\
  H532       & \phnnn1.484698     & 0.0035    & 267\phdn & 48170.37\phn & 40.31\phn & 68.94    & $\phn94\phdnn\pm15\phdnn$    & $0.73\pm0.12$ &  $\phnn1.4075\pm\phn0.0044$       & $\phnnn1.92\pm\phnn0.30$       & $0.38\phn\pm0.13\phn$ & $0.281\phn\pm0.052\phn$\\
  vB 96      & 5100\phdnnnnnn     & 0.664\phn & 310.4    & 50106\phdnnn & 41.14\phn & \phn4.77 & $\phnn9.6\phn\pm\phn3.5\phn$ & $0.50\pm0.18$ &  $249.8\phnnn\pm\phn8.1\phnnn$    & $\phn500\phdnn\pm180\phdnn$    & $0.44\phn\pm0.37\phn$ & $0.22\phnn\pm0.11\phnn$\\
  L79        & 3688\phdnnnnnn     & 0.765\phn & 341.8    & 48662.1\phnn & 41.454    & \phn4.31 & $\phn27\phdnn\pm19\phdnn$    & $0.16\pm0.11$ &  $140.8\phnnn\pm\phn7.9\phnnn$    & $\phn890\phdnn\pm620\phdnn$    & $2.7\phnn\pm5.2\phnn$ & $0.43\phnn\pm0.53\phnn$\\
  vB 102     & \phn734.79\phnnnn  & 0.513\phn & 337.2    & 50015.8\phnn & 40.192    & \phn3.92 & $\phn31.1\phn\pm\phn5.9\phn$ & $0.13\pm0.02$ &  $\phn34.0\phnnn\pm\phn1.4\phnnn$ & $\phn270\phdnn\pm\phn51\phdnn$ & $1.84\phn\pm0.97\phn$ & $0.231\phn\pm0.080\phn$\\
  vB 142     & \phn975.7\phnnnnn  & 0.675\phn & 271.9    & 50383.6\phnn & 41.684    & \phn1.19 & $\phn21\phdnn\pm14\phdnn$    & $0.06\pm0.04$ &  $\phn11.8\phnnn\pm\phn1.3\phnnn$ & $\phn220\phdnn\pm140\phdnn$    & $0.42\phn\pm0.82\phn$ & $0.024\phn\pm0.031\phn$\\
  vB 113     & 2429\phdnnnnnn     & 0.327\phn & 294.1    & 51404\phdnnn & 41.544    & \phn3.36 & $\phn11.1\phn\pm\phn2.9\phn$ & $0.30\pm0.08$ &  $106.1\phnnn\pm\phn4.3\phnnn$    & $\phn350\phdnn\pm\phn91\phdnn$ & $0.49\phn\pm0.32\phn$ & $0.150\phn\pm0.060\phn$\\
  vB 115     & 1208.2\phnnnnn     & 0.480\phn & 129.1    & 50731.6\phnn & 40.730    & \phn5.65 & $\phn10.9\phn\pm\phn3.7\phn$ & $0.52\pm0.18$ &  $\phn82.3\phnnn\pm\phn1.5\phnnn$ & $\phn158\phdnn\pm\phn54\phdnn$ & $0.25\phn\pm0.20\phn$ & $0.131\phn\pm0.059\phn$\\
  vB 121     & \phnnn5.750872     & 0.361\phn & \phn42.0 & 49743.257    & 41.28\phn & 20.11    & $122.4\phn\pm\phn9.2\phn$    & $0.16\pm0.02$ &  $\phnn1.483\phn\pm\phn0.015\phn$ & $\phnnn9.03\pm\phnn0.68$       & $1.20\phn\pm0.25\phn$ & $0.197\phn\pm0.026\phn$\\
  vB 151     & \phn629.37\phnnnn  & 0.297\phn & 203.8    & 50544.0\phnn & 42.167    & \phn5.22 & $\phnn9.8\phn\pm\phn2.2\phn$ & $0.53\pm0.12$ &  $\phn43.14\phnn\pm\phn0.89\phnn$ & $\phnn81\phdnn\pm\phn18\phdnn$ & $0.125\pm0.065$       & $0.067\phn\pm0.020\phn$\\
  BD+02 1102 & \phnn32.5121\phnn  & 0.344\phn & 228.7    & 51056.33\phn & 47.60\phn & 21.31    & $\phn53\phdnn\pm21\phdnn$    & $0.40\pm0.15$ &  $\phnn8.94\phnn\pm\phn0.21\phnn$ & $\phnn22.4\phn\pm\phnn8.6\phn$ & $0.82\phn\pm0.78\phn$ & $0.33\phnn\pm0.19\phnn$\\
  \enddata
\end{deluxetable}

\begin{deluxetable}{lccccccccc}
  \tablecaption{Alternate Orbital Solutions\label{table:altorbits}}
  \tablewidth{0pt}
  \tablecolumns{10}
  \tabletypesize{\footnotesize}
  \setlength{\tabcolsep}{0.025in}
  \tablehead{
    \colhead{} & 
    \colhead{\period} & 
    \colhead{} &
    \colhead{\w} &
    \colhead{\Tknot} &
    \colhead{\g} & 
    \colhead{\Kone} & 
    \colhead{\Ktwo} &
    \colhead{} &
    \colhead{} \\
      
    \colhead{Target} & \colhead{(days)} & \colhead{\e} &
    \colhead{(deg)} & \colhead{(JD-2400000)} & \colhead{(\kps)} &
    \colhead{(\kps)} & \colhead{(\kps)} & \colhead{\q} & \colhead{SB1 Ref.}} 
    \startdata
  L20        & \phnnn2.394357     & 0.057 & 279.2       & 43892.36    & 36.76    & 66.2\phn & $\phn98.8\pm\phn2.7$     & $0.67\pm0.02$ & G81 \\
  vB 40      & \phnnn4.00050\phn  & 0.060 & \phn13\phdn & 22274.81    & 37.4\phn & 36.1\phn & $\phn88.2\pm\phn2.7$     & $0.41\pm0.02$ & S21 \\
  vB 43      & \phn590.6\phnnnnn  & 0.638 & 303.1       & 43512.9\phn & 39.81    & \phn9.91 & $\phn14.1\pm\phn1.3$     & $0.70\pm0.07$ & G85 \\
  vB 62      & \phnnn8.55089\phn  & 0.233 & \phn38.0    & 42588.22    & 38.77    & 16.46    & $\phn87\phdn\pm30\phdn$  & $0.19\pm0.06$ & G78 \\
  L57        & 1907\phd\phnnnnnn  & 0.485 & \phn95\phdn & 43470.5\phn & 40.23    & \phn6.83 & $\phnn6.8\pm\phn2.2$     & $1.00\pm0.32$ & G85 \\
  vB 69      & \phnn41.6625\phnn  & 0.662 & 326.9       & 43650.67    & 39.81    & \phn7.28 & $\phn69\phdn\pm17\phdn$  & $0.10\pm0.03$ & G85 \\
  vB 77      & \phn238.87\phnnnn  & 0.242 & 127\phdn    & 43298\phdnn & 39.81    & \phn6.53 & $\phn20.1\pm\phn1.9$     & $0.32\pm0.03$ & G85 \\
  H509       & \phn844.6\phnnnnn  & 0.148 & 325\phdn    & 44413\phdnn & 40.32    & \phn6.20 & $\phn12.6\pm\phn1.6$     & $0.49\pm0.06$ & G85 \\
  vB 121     & \phnnn5.75096\phn  & 0.354 & \phn54.9    & 42192.06    & 42.74    & 19.70    & $126.8\pm\phn8.1$        & $0.16\pm0.01$ & G78 \\
  \enddata
  \tablerefs{G78 -- \citet{griffin1978}; G81 -- \citet{griffin1981}; G85 -- \citet{griffin1985}; S21 -- \citet{sanford1921}}
\end{deluxetable}

\begin{deluxetable}{lcc}
  \tablecaption{SB2 Component Masses\label{table:componentmasses}}
  \tablewidth{0pt}
  \tablecolumns{3}
  \tabletypesize{\footnotesize}
  \tablehead{\colhead{Target} & \colhead{\Mone (\msun)} & \colhead{\Mtwo (\msun)}}
  \startdata
  vB 9      & $1.15\pm0.05$ & $0.41\pm0.06$ \\     
  L20       & $0.88\pm0.04$ & $0.60\pm0.04$ \\
  H69\tablenotemark{a}   & $0.45\pm0.10$ & $0.27\pm0.06$ \\
  L33       & $0.93\pm0.05$ & $0.64\pm0.06$ \\
  vB 40     & $1.38\pm0.10$ & $0.63\pm0.05$ \\
  vB 43     & $1.01\pm0.06$ & $0.67\pm0.06$ \\
  vB 59\tablenotemark{b} & $0.95\pm0.08$ & $0.17\pm0.06$ \\
  vB 62     & $1.23\pm0.03$ & $0.28\pm0.04$ \\
  H382      & $1.09\pm0.04$ & $0.40\pm0.06$ \\
  L57\tablenotemark{b}   & $0.77\pm0.09$ & $0.66\pm0.19$ \\
  vB 68\tablenotemark{c} & $1.70\pm0.08$ & $1.19\pm0.20$ \\
  vB 69     & $1.03\pm0.04$ & $0.16\pm0.02$ \\
  H441\tablenotemark{a}  & $0.61\pm0.10$ & $0.27\pm0.05$ \\      
  vB 77     & $1.29\pm0.03$ & $0.43\pm0.04$ \\
  H509      & $0.83\pm0.04$ & $0.38\pm0.12$ \\
  H532      & $0.79\pm0.08$ & $0.58\pm0.15$ \\
  vB 96     & $1.00\pm0.05$ & $0.49\pm0.18$ \\
  L79\tablenotemark{a}   & $0.61\pm0.10$ & $0.10\pm0.07$ \\
  vB 102    & $1.19\pm0.04$ & $0.15\pm0.04$ \\
  vB 142    & $1.10\pm0.04$ & $0.07\pm0.04$ \\
  vB 113    & $1.17\pm0.03$ & $0.32\pm0.12$ \\
  vB 115    & $0.97\pm0.05$ & $0.50\pm0.18$ \\
  vB 121    & $1.31\pm0.04$ & $0.21\pm0.03$ \\
  vB 151    & $1.03\pm0.09$ & $0.55\pm0.13$ \\
  BD+02 1102& $1.29\pm0.05$ & $0.52\pm0.20$ \\
  \enddata
  \tablenotetext{a}{Estimated from primary spectral type}
  \tablenotetext{b}{Tycho parallax}
  \tablenotetext{c}{\citet{pinsonneault2004} isochrone}
\end{deluxetable}     

\clearpage

\end{document}